\documentclass[namedreferences]{solarphysics}

\usepackage[hyperref,optionalrh,showbiblabels]{spr-sola-addons} 
\usepackage{graphicx}        
\usepackage{gensymb}
\usepackage{color}           
\newcommand\arcsec{\mbox{$^{\prime\prime}$}}%

\chardef\us=`\_

\begin{document}

\begin{article}
\begin{opening}

\title{The Sunspot Solar Observatory Data Archive: Continuing Operations at the Dunn Solar Telescope\\ {\it Solar Physics}}

\author[addressref=nmsu, corref, email={sellers@nmsu.edu}]{\inits{S.G.~}\fnm{Sean~G.~}\lnm{Sellers}\orcid{0000-0001-5342-0701}}
\address[id={nmsu}]{Department of Astronomy, New Mexico State University, New Mexico, USA}

\author[addressref=nmsu]{\inits{J.~}\fnm{Juie~}\lnm{Shetye}\orcid{0000-0002-4188-7010}}

\author[addressref=csun]{\inits{D.J.~}\fnm{Damian~J.~}\lnm{Christian}\orcid{0000-0003-1746-3020}}
\address[id={csun}]{Department of Physics and Astronomy, California State University Northridge, 18111 Nordhoff St, Northridge, CA 91330, USA}

\author[addressref={qub}]{\inits{D.B.~}
\fnm{David~B.~}
\lnm{Jess}
\orcid{0000-0002-9155-8039}}
\address[id={qub}]{Astrophysics Research Centre, School of Mathematics and Physics, Queen's University Belfast, Belfast, BT7 1NN, UK}

\author[addressref={qub}]{\inits{P.H.~}\fnm{Peter~H.~}\lnm{Keys}\orcid{0000-0001-8556-470X}}

\author[addressref={nmsu,apo}]{\inits{G.A.~}\fnm{Gordon~A.~}\lnm{MacDonald}\orcid{0000-0002-5259-5507}}
\address[id={apo}]{Apache Point Observatory, Sunspot, NM, 88349}

\author[addressref={nmsu}]{\inits{R.T.J~}\fnm{R.T.~James~}\lnm{McAteer}\orcid{0000-0003-1493-101X}}

\author[addressref={nmsu}]{\inits{J.~}\fnm{Jason~}\lnm{Jackiewicz}\orcid{0000-0001-9659-7486}}

\author[addressref={nmsu}]{\inits{C.~}\fnm{Colin~}\lnm{Hancock}\orcid{0009-0004-4810-1051}}
\author[addressref={nmsu}]{\inits{M.S.~}\fnm{Michael~S.~}\lnm{Thompson}}
\author[addressref={apo}]{\inits{J.E.~}\fnm{Jamey~E.~}\lnm{Eriksen}\orcid{0009-0004-8297-2272}}
\author[addressref={nmsu}]{\inits{S.~}\fnm{Sara~}\lnm{Jeffreys}\orcid{0009-0009-9085-159X}}

\runningauthor{Sunspot Solar Observatory Consortium}
\runningtitle{SSODA Archive}

\begin{abstract}
The Sunspot Solar Observatory Data Archive (SSODA) stores data acquired with the suite of instruments at the Richard B. Dunn Solar Telescope (DST) from February 2018 to the present. 
The instrumentation at the DST continues to provide high cadence imaging, spectroscopy, and polarimetry of the solar photosphere and chromosphere across a wavelength range from 3500~{\AA} to 11{\,}000~{\AA}. 
At time of writing, the archive contains approximately 374~TiB of data across more than 520 observing days (starting on February 1, 2018). 
These numbers are approximate as the DST remains operational, and is actively adding new data to the archive. 
The SSODA includes both raw and calibrated data. 
A subset of the archive contains the results of photospheric and chromospheric spectropolarimetric inversions using the Hazel-2.0 code to obtain maps of magnetic fields, temperatures, and velocity flows. 
The SSODA represents a unique resource for the investigation of plasma processes throughout the solar atmosphere, the origin of space weather events, and the properties of active regions throughout the rise of Solar Cycle 25.  
\end{abstract}
\end{opening}

\section{Introduction}
     \label{S-Introduction} 

The solar atmosphere is a complex and dynamic environment, rife with diverse structures produced by the continual dance along the edge of magnetically- and thermally-dominant plasmas. 
These structures, despite occurring at extremely small temporal and spatial scales, influence events across the entire heliosphere \citep[e.g.,][]{Parker1957, Sweet1958, 2019LRSP...16....5V}.
Observations and computational modeling undertaken throughout the last quarter-century have made significant leaps towards characterizing the interactions of solar plasmas, their formations, dynamics, and oscillatory nature \citep[e.g.,][and references therein]{Muram, Rempel2009, bifrost, Morton2012, MM2013, Rempel2014, Borrero2017, Casini2017, Jess2020, 2023LRSP...20....1J}. 
Many of these groundbreaking advancements were facilitated using the Dunn Solar Telescope \citep[DST;][]{DST}, which continues to cement its legacy as a world-class facility with robust, high-quality scientific results, e.g. recent results described in \citet{Jess2015, 2017ApJ...842...59J}, \citet{2015ApJ...806..132G, Grant2018, 2022ApJ...938..143G}, \citet{Christian2019}, \citet{2021RSPTA.37900172G}, \citet{Bate2022, 2024ApJ...970...66B}, \citet{2022NatCo..13..479S}, \citet{Vesa2023}, and \citet{2024PhRvL.132u5201M}.
\par
Since 2017, the DST has operated under the control of the Sunspot Solar Observatory Consortium (SSOC), members of which currently include New Mexico State University (NMSU, operating partner), the National Solar Observatory (NSO), the High Altitude Observatory (HAO), the University of Hawai'i Institute for Astronomy (IFA), Queen's University Belfast (QUB), California State University Northridge (CSUN), and the University of Colorado Boulder (CU Boulder). 
During consortium operations, a selection of up to four facility instruments are operated simultaneously, and targets are often chosen to complement those observed by various space-borne telescopes, such as the Hinode Solar Optical Telescope \citep[Hinode/SOT;][]{SOT}, the Interface Region Imaging Spectrograph \citep[IRIS;][]{IRIS}, the Solar Dynamics Observatory Helioseismic and Magnetic Imager and Atmospheric Imaging Assembly \citep[SDO/HMI and SDO/AIA;][respectively]{HMI, AIA}, as well as targets of interest for in situ facilities such as the Parker Solar Probe \citep[PSP;][]{PSP1,PSP2}.
\par
  The SSOC has also implemented several key science projects in addition to daily operations. 
  One key project is to acquire diffraction-limited imaging, spectropolarimetric, and spectroscopic observations spanning the photosphere, through the chromosphere to the outer corona in support of the PSP perihelia passages. 
  The PSP makes in situ measurements to answer fundamental questions about the coronal conditions leading to the nascent solar wind and eruptive transients that create space weather.  
  The origins and acceleration mechanisms of the slow solar wind remain mysterious, as do the strong fluctuations and non-Maxwellian velocity distributions observed in the fast solar wind (see the review by \citealt{Dudik2017} and the simulated PSP encounters from \citealt{Roberts2018}).
  Having ground-based observations to complement the PSP in situ measurements is critical to answer these key questions \citep[e.g.,][]{SWEAP2016, Gombosi2018}. 
  Other key projects include observations of the quiet Sun, active regions, and  filaments.  The high spatial- and temporal-resolution observations available with the DST suite of instruments will help us gain a better understanding of the rapid variability in the solar magnetic field under many different conditions. 
  The large field of view (FOV) of the DST imaging instruments is also very complementary to the smaller FOV of larger telescopes, such as DKIST.
\par
The goal of this work is to highlight recent efforts undertaken in archiving, characterizing, and disseminating data collected at the DST under consortium operations via the Sunspot Solar Observatory Data Archive (SSODA) portal. 
These data are publicly available, and the archive contains significant quantities of reduced, science-ready data. 
The data products, instruments, processing levels, and science products are defined in Section~\ref{sec:instr}. 
Section~\ref{sec:obs_camp} describes the common synoptic observational campaigns carried out at the DST, as well as an outline of the amount and characteristics of data archived for each of these campaigns. 
Significant efforts have been made to characterize the quality of the obtained data, using data- and telescope-specific algorithms, described in Section~\ref{sec:quality}. 
An outline of data access, as well as currently-tracked metadata is presented in Section~\ref{sec:metadata}.
An example of high-quality, but scientifically-unexplored data is presented in Section~\ref{sec:usecase}, and finally, a summary of the work can be found in Section~\ref{sec:conclusions}.

\section{Instrumental Complement} 
      \label{sec:instr}
The SSODA contains data from five instruments, from which a subset may be chosen for daily observations. 
Each instrument may have vastly different data products, with correspondingly different data level definitions. 
Most data are stored in the Flexible Image Transport System (FITS) format, however, legacy data products may be stored in other formats, such as binary, or save objects from Interactive Data Language (IDL) pipelines. 
Software to read these data products is available either in commonly-accessed Python data repositories (e.g., astropy.io.fits and scipy.io.readsav), or can be read via SSOC-maintained code repositories. 
The following sections contain a basic summary of the telescope system, followed by a description of the instrument systems and the typically-available data products for each instrument. 

\subsection{The Dunn Solar Telescope}\label{sec:instr.dst}
The DST is an evacuated solar telescope located at an altitude of $2{\,}800~{\mathrm{m}}$ at an observing site in southern New Mexico, USA. 
The telescope features a $76~{\mathrm{cm}}$ fused quartz entrance window positioned $40~{\mathrm{m}}$ above the ground level. 
This entrance window sets both the diffraction limit of the system, and by its composition, the available spectral range of the system. 
The typical instrumental complement supports studies from 3500~{\AA} to 11{\,}000~{\AA}, however, the telescope retains the ability to study wavelengths down to $\approx2.5~{\mathrm{\mu m}}$, should a future instrument be commissioned to probe these wavelengths. 
Across this wavelength range, the diffraction-limited resolution of the telescope is between $0{\,}.{\!\!}''116$ (at 3500~{\AA}) and $0{\,}.{\!\!}''363$ (at 11{\,}000~{\AA}).
\par
The primary mirror of the telescope is located $54~{\mathrm{m}}$ below ground level. 
The primary mirror is a spherical segment, with a $1.5~{\mathrm{m}}$ diameter, and a $54.864~{\mathrm{m}}$ focal length. 
The prime focal plane has a plate scale of $3{\,}.{\!\!}''76$ per mm and a focal ratio, \textit{f}-72. 
The field stop at this position is $45~{\mathrm{mm}}$ square, giving the DST a maximum FOV of $168{\,}.{\!\!}''75$ square. 
The high focal ratio of the system serves to minimize the spherical aberration induced by the geometry of the primary mirror.
\par
The DST is equipped with a high-order adaptive optics system \citep[referred to in technical documentation as HOAO or AO-76;][]{RimmeleLOAO, RimmeleHOAO}, consisting of a fast tip-tilt mirror, a 76-subaperture Shack-Hartmann wavefront sensor system, and a 97-segment deformable mirror. 
Together, these systems operate with a loop update rate of 2.5~kHz, allowing the DST to observe structures near the diffraction limit, with the fast-imaging systems able to recover structures at the diffraction limit with additional processing techniques (detailed in Section~\ref{sec:instr.rosa}).

\subsection{Facility InfraRed Spectropolarimeter} 
\label{sec:instr.firs}
The Facility InfraRed Spectrpolarimeter \citep[FIRS;][]{FIRS}, is a dual-beam rastering slit spectropolarimeter with cotemporal slit-jaw imaging, operated under consortium direction from 2017 to the present.
While the instrument is configured to provide full-Stokes spectropolarimetry of several lines, in consortium use, the instrument is typically configured to observe the He~{\sc{i}} 10830~{\AA} triplet and nearby photospheric lines. 
The pixel scale of this instrument during normal consortium operations is $0{\,}.{\!\!}''15$ per pixel along the slit for a slit length of $\approx75${\arcsec} with a slit width of $0{\,}.{\!\!}''3$. 
Rastering is typically done in a dense fashion, with each slit position contiguous to the previous. 
Exceptions are rare, and noted in both the daily observing logs and in the Level-1.5  FITS headers. 
The maximum FOV for a single raster is limited by the telescope's field stop, and corresponds to a 550-step raster (165\arcsec). 
The linear spectral dispersion in the He~{\sc{i}} 10830~{\AA} region is 0.038~{\AA} per pixel (with a resolution twice this value, corresponding to a spectral resolving power, $R = 142{\,}500$) between $10819~{\mathrm{\AA}} \lesssim \lambda \lesssim 10858~{\mathrm{\AA}}$. 
FIRS is typically operated with a cotemporal slit-jaw imager to allow for precise coalignment between the spectropolarimeter and other instruments. The slit-jaw imager is typically operated with a $6300$~\AA\ broadband filter. 
FIRS data are available within the SSODA as Level-0, Level-1, Level-1.5, and Level-2 data products. 
These data levels correspond to the following data products:
\begin{itemize}
    \item \textbf{Level-0}: FIRS Level-0 data are the raw data obtained from the telescope. 
    These data are in FITS format, with each file corresponding to a single slit position, containing eight polarization modulation states. 
    \item \textbf{Level-1}: FIRS Level-1 data are reduced via the standard IDL \texttt{firs-soft} package. Details of this and all other relevant reduction codes can be found in the code availability statement. 
    This pipeline corrects for dark current, flat-fielding, and performs a polarimetric calibration including \textit{I} $\rightarrow$ \textit{Q}, \textit{U}, \textit{V} crosstalk. 
    Data at level one are stored as binary, with an IDL save file providing metadata from the calibration process.
    \item \textbf{Level-1.5}: FIRS data at Level-1.5 are subjected to further processing using a set of Python-based tools designed to remove lingering artifacts from the \texttt{firs-soft} reduction pipeline. 
    Level-1.5 data are wavelength calibrated, and corrected for fringes, spectral transmission, spectral deformations along the slit, and \textit{V}~$\rightarrow$~\textit{Q},~\textit{U} crosstalk\footnote{\textit{V}$\rightarrow$\textit{Q},~\textit{U} crosstalk is corrected using the assumption that solar structures appear different between integrated \textit{V}, \textit{Q}, and \textit{U}. For datasets with little to no structure, this correction is not performed, and \textit{V}~$\rightarrow$~\textit{Q},~\textit{U} crosstalk is assumed to be negligible}. 
    Data at Level-1.5 are repacked to FITS format, similar in structure to the \textit{IRIS} Level-3 data product, with separate extensions for \textit{I}, \textit{Q}, \textit{U}, and \textit{V} measurements, an extension for the timestamps of each slit position, and an extension containing the wavelengths as determined from comparison to the Fourier Transform Spectrograph (FTS) atlas \citep{FTS1, FTS2}. Level-1.5 data files are accompanied by map files containing derived parameters for each of the Si~{\sc{i}}~10827~{\AA}, He~{\sc{i}}~10829~{\AA}, and blended He~{\sc{i}}~10830~{\AA} lines. Derived parameters include total, mean, and net circular polarization, mean linear polarization, and both line-of-sight (LOS) Doppler velocities and Doppler velocity widths from moment analysis techniques.  
    \item \textbf{Level-2}: FIRS data at Level-2 have been inverted using the Hazel 2.0 \citep{hazel} code, with varying atmospheric structures and initial conditions. 
    Depending on the target, Level-2 files may contain inversions of the Si~{\sc{i}} line using the Stokes Inversion based on Response functions \citep[SIR][]{SIR} code wrapper within the Hazel code, as well as the He~{\sc{i}} line. 
    Level-2 data are stored in FITS format, with separate extensions for photospheric and chromospheric inversion results. 
    Where both are present, further extensions contain the fit Stokes profiles, as well as the normalized profiles provided to the Hazel code.
    Hazel+SIR inversion results provide the full (non-disambiguated) magnetic field vector (expressed as X, Y, and Z components), LOS velocity, and velocity width for both the photosphere and chromosphere. Photospheric inversions provide velocity width in terms of micro- and macro-turbulent velocities, and an estimate of temperature at a range of optical depths. Chromospheric inversions assume a slab chromospheric model, and do not have an optical depth range, rather, optical depth is fit for, along with plasma-$\beta$ and the chromospheric damping parameter, $a$. While chromospheric temperature is not fit for explicit use in Hazel, the plasma-$\beta$, being a ratio between magnetic and thermal pressure, can be used as a proxy.
\end{itemize}

\begin{figure}
    \centering
    \includegraphics[width=\textwidth]{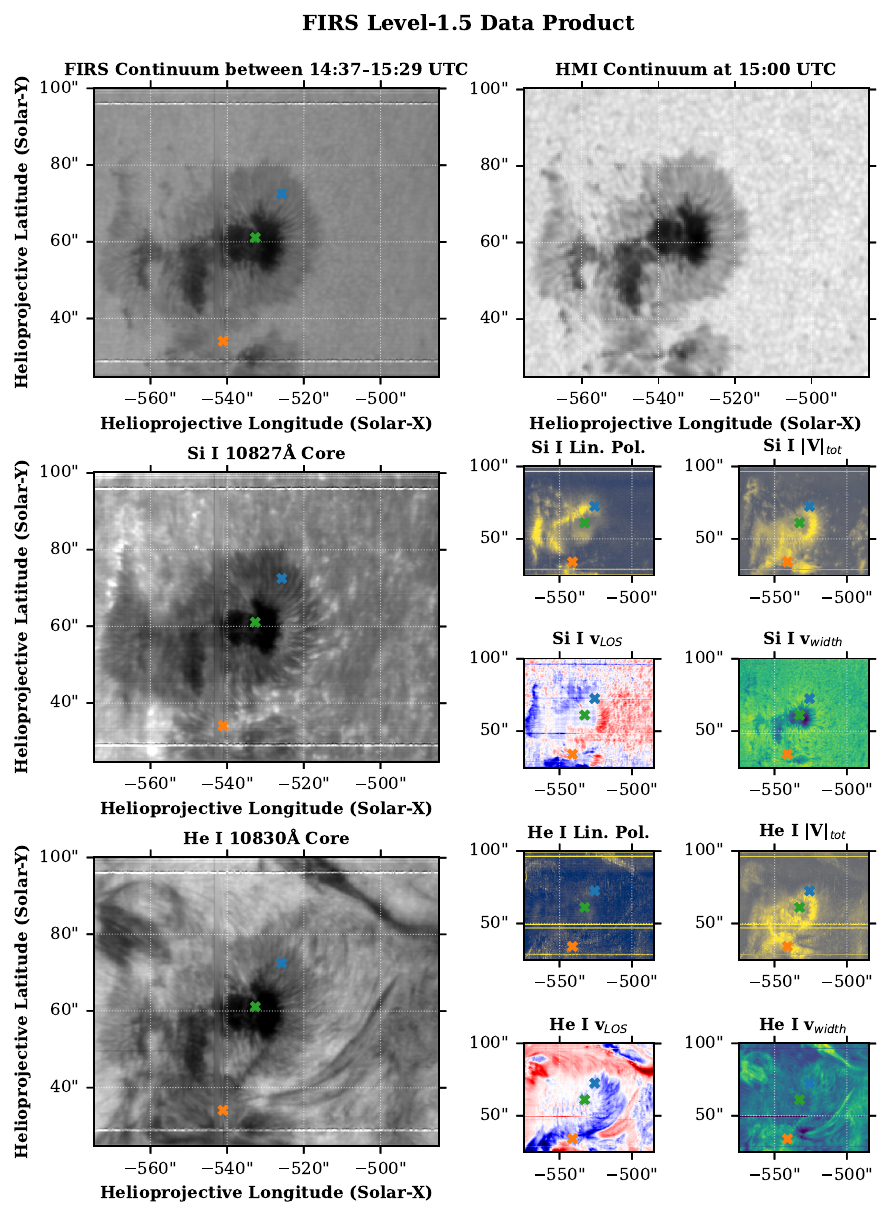}
    \caption{Examples of FIRS Level-1.5 maps from an active region dataset obtained on September 20, 2023. The top row contains an image of continuum intensity, as observed by the FIRS instrument (left) compared to the same region as observed in HMI continuum intensity. The horizontal lines running across the FIRS FOV are hairline fiducials, used for co-aligning FIRS spectral data. The middle row contains observations of the Si~{\sc{i}}~10827~{\AA} line, in line-core intensity, total linear and circular polarization (units of integrated DN), line-of-sight velocity (km~s$^{-1}$), and line width (km~s$^{-1}$). The bottom row contains the same observations for the He~{\sc{i}}~10830~{\AA} line. The marked points denote the locations of the spectra shown in Figure~\ref{fig:firs-spex}.}
    \label{fig:firs-l1.5}
\end{figure}

\begin{figure}
    \centering
    \includegraphics[width=\textwidth]{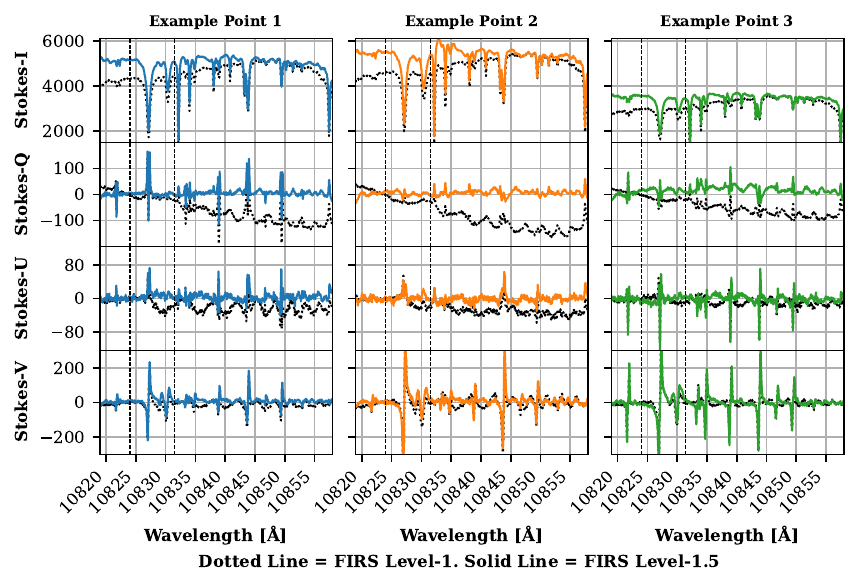}
    \includegraphics[width=\textwidth]{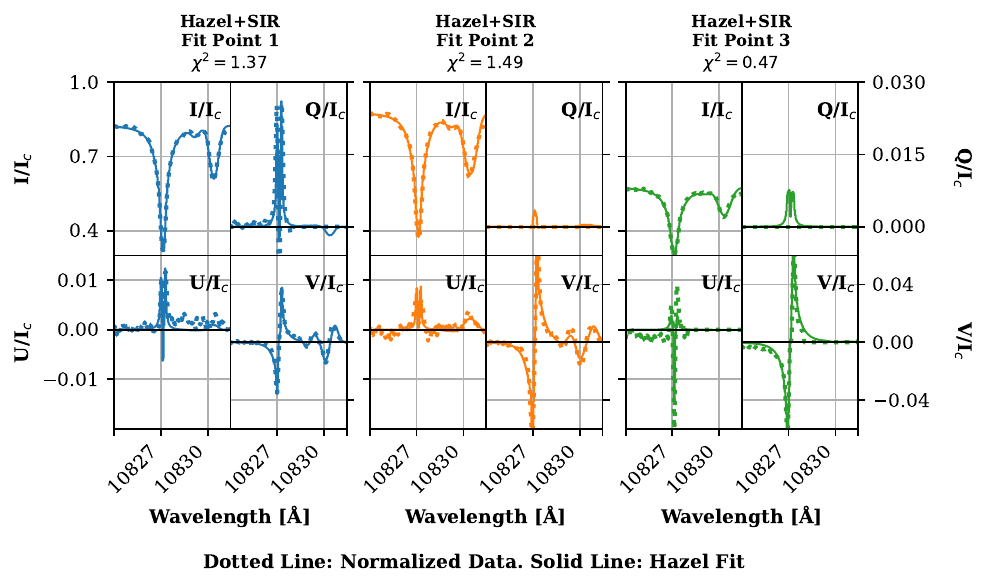}
    \caption{
    \textbf{Top:} Example of FIRS Level-1 and 1.5 Stokes spectra from data obtained on September 20, 2023. Level-1 data are shown as dotted lines, while Level-1.5 are shown as solid lines for each, \textit{I}, \textit{Q}, \textit{U}, and \textit{V}. This comparison highlights the additional corrections employed in the formation of the Level-1.5 data product.
    \textbf{Bottom:} Example of FIRS Level-2 spectra, showcasing the normalization used in pre-processing for the Hazel code (dotted line), alongside the Hazel synthesized spectra (solid line). Each column in both graphics corresponds to the locations shown in Figure~\ref{fig:firs-l1.5}.
    }
    \label{fig:firs-spex}
\end{figure}

\begin{figure}
    \centering
    \includegraphics[width=\textwidth]{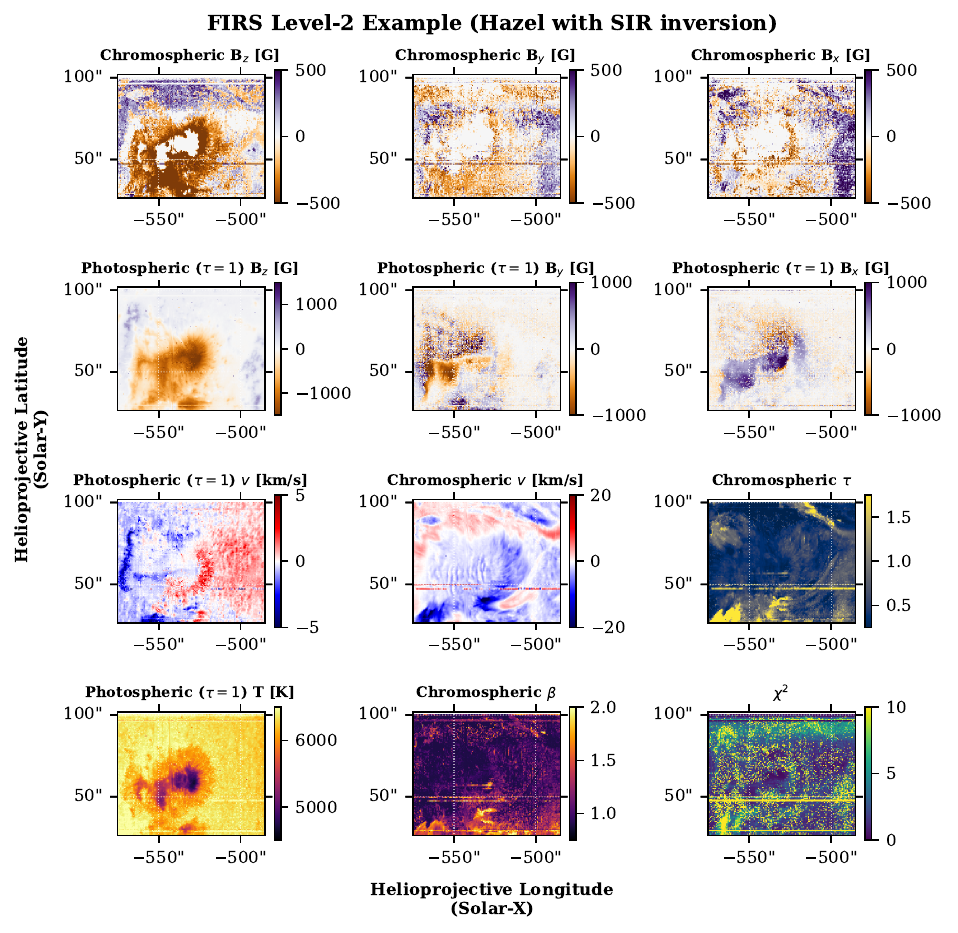}
    \caption{Physical parameters obtained from Hazel fitting of FIRS spectropolarimetry from data obtained on September 20, 2023. The top row contains the components of the magnetic vector, \textbf{B}, for the He{~\sc{i}} line. The second row contains the magnetic vector for the photospheric Si{~\sc{i}} line. No attempt is made to resolve the 180-degree ambiguity in the FIRS Level-2 data product at this time, and thus the B$_x$ and B$_y$ components are best understood as the pseudo-vector. The third row contains, left-to-right, the photospheric LOS velocity, chromospheric LOS velocity, and the optical depth of chromospheric plasma, $\tau$. The bottom row contains, left-to-right, the photospheric temperature at the ``surface'' (an optical depth, $\tau$, of unity), the chromospheric plasma-$\beta$, and the reduced $\chi^2$ as a goodness-of-fit measure. These parameters form part of the FIRS Level-2 data product, and this particular example was created using the Level-1.5 data shown in Figures~\ref{fig:firs-l1.5}~and~\ref{fig:firs-spex}, and the same three data points are marked on each panel. The presence of horizontal stripes in the data is indicative of the presence of the spectrograph fiducial ``hairlines'' in the case of those stripes near the top and bottom of each image. Other artifacts are caused by partially corrected hot pixels on the spectrograph chip.}
    \label{fig:firs-hazel}
\end{figure}

Examples of FIRS data can be seen in Figures~\ref{fig:firs-l1.5},~\ref{fig:firs-spex},~and~\ref{fig:firs-hazel}. 
Figure~\ref{fig:firs-l1.5} shows maps of FIRS Level-1.5 data, including continuum and line-core intensity, total degree of linear and circular polarization, line-of-sight velocity, and width for the Si~{\sc{i}} photospheric and He~{\sc{i}} chromospheric lines. 
The locations marked with an ``x'' correspond to the data shown in Figure~\ref{fig:firs-spex}. Figure~\ref{fig:firs-spex} shows progressive examples of FIRS data from Level-1 to Level-1.5, and finally to Level-2, demonstrating successive corrective techniques. 
The Level-1.5 data (solid lines) are shown compared to Level-1 data in the top half of the figure, while the bottom half of the figure shows these points, now windowed to include only the Si~{\sc{i}}{\,}$-${\,}He~{\sc{i}} complex, and normalized to the quiet-Sun continuum intensity as determined by local quiet-Sun intensity, corrected for heliocentric position angle (dotted line), along with the spectral fits provided by Hazel inversions (solid line). 
Figure~\ref{fig:firs-hazel} shows the net result of Hazel fitting; maps of various photospheric and chromospheric physical parameters. 

\subsection{Rapid Oscillations in the Solar Atmosphere and H$\alpha$ Rapid Dynamics CAMera}\label{sec:instr.rosa}
The Rapid Oscillations in the Solar Atmosphere \citep[ROSA;][]{ROSA} system, and the associated H$\alpha$ Rapid Dynamics CAMera \citep[HARDcam;][]{2012ApJ...757..160J} system, are an array of cameras operated in concert as a rapid imaging instrument. 
The system has been operated under consortium direction from 2017 to the present. 
The ROSA instrument consists of up to six (though under consortium operations, only up to four are used) externally triggered and synchronized cameras, operating primarily in wavelengths $\lambda <$5000~{\AA} . 
The spatial resolution of these cameras can be altered for the purpose of acquiring either diffraction-limited images, or studying a wide FOV. 
When operated at the diffraction limit, the ROSA cameras observe a 60{\arcsec} square FOV with a plate scale of $0{\,}.{\!\!}''06$ per pixel, and a maximal spatial resolution after reconstruction of $0{\,}.{\!\!}''13$ at $4000$~{\AA}. 
When operated for a wider FOV, the ROSA cameras observe a 156{\arcsec} square field, with a plate scale of $0{\,}.{\!\!}''156$ per pixel. 
The HARDcam system is typically operated with a plate scale of $0{\,}.{\!\!}''0846$ per pixel and observes a 173{\arcsec} square field, with a maximum spatial resolution of $0{\,}.{\!\!}''254$ at wavelength of H$\alpha$ (6563~{\AA}), just above the diffraction limit for the system. Unless otherwise noted, the cameras are operated in the configuration given in Table~\ref{tab:rosa}.

\begin{table}
\caption{Typical optical setup for consortium observations using ROSA/HARDcam includes ROSA cameras 1\textendash 3 and HARDcam. ROSA camera 4 is used infrequently. 
}
\label{tab:rosa}
\begin{tabular}{llcccc}     
  \hline                   
Camera Name & Filter & Wavelength & FWHM & Typ. Framerate & No. Pixels \\
            &        & \AA        & \AA  & Hz \\
  \hline
ROSA DAS1 & G-Band & 4305 & 9.2 & 30 & 1002$\times$1004\\
ROSA DAS2 & Continuum & 4170 & 52.0 & 30 & 1002$\times$1004\\
ROSA DAS3 & Ca~{\sc{ii}}~K & 3934 & 1.2 & 6, 15, or 30 & 1002$\times$1004\\
ROSA DAS4 & 3500 & 3500 & 102.0 & 1\textendash6 & 1002$\times$1004\\
\hline
HARDcam Zyla & Zeiss H$\alpha$ & 6562.8 & 0.25 & 29.4 & 2048$\times$2048\\
\hline
\end{tabular}
\end{table}

ROSA and HARDcam data are available within the SSODA as Level-0, Level-1, and Level-1.5 data products. 
These data levels correspond to the following data products:
\begin{itemize}
    \item \textbf{Level-0:} ROSA/HARDcam Level-0 data are the raw imaging data as obtained by the instrument control computers. 
    ROSA data are stored in FITS format, with up to 256 image extensions per file. HARDcam data are stored as binary images. 
    ROSA and HARDcam Level-0 data may be compressed using the FITS-specific \texttt{fpack} algorithm.
    \item \textbf{Level-1:} ROSA/HARDcam Level-1 data have been subjected to dark-correction and flat-fielding procedures, using the standard SSOsoft pipeline, and have been processed using the KISIP speckle-burst reconstruction algorithm \citep{kisip1, kisip2}. 
    Level-1 data are stored in FITS format, with a single speckle-burst reconstructed image stored per file.
    \item \textbf{Level-1.5:} ROSA/HARDcam Level-1 data have been destretched using an iterative subfield-based spline destretch algorithm, with an optional lateral solar flow-preserving subroutine.
\end{itemize}

\begin{figure}
    \centering
    \includegraphics[width=\textwidth]{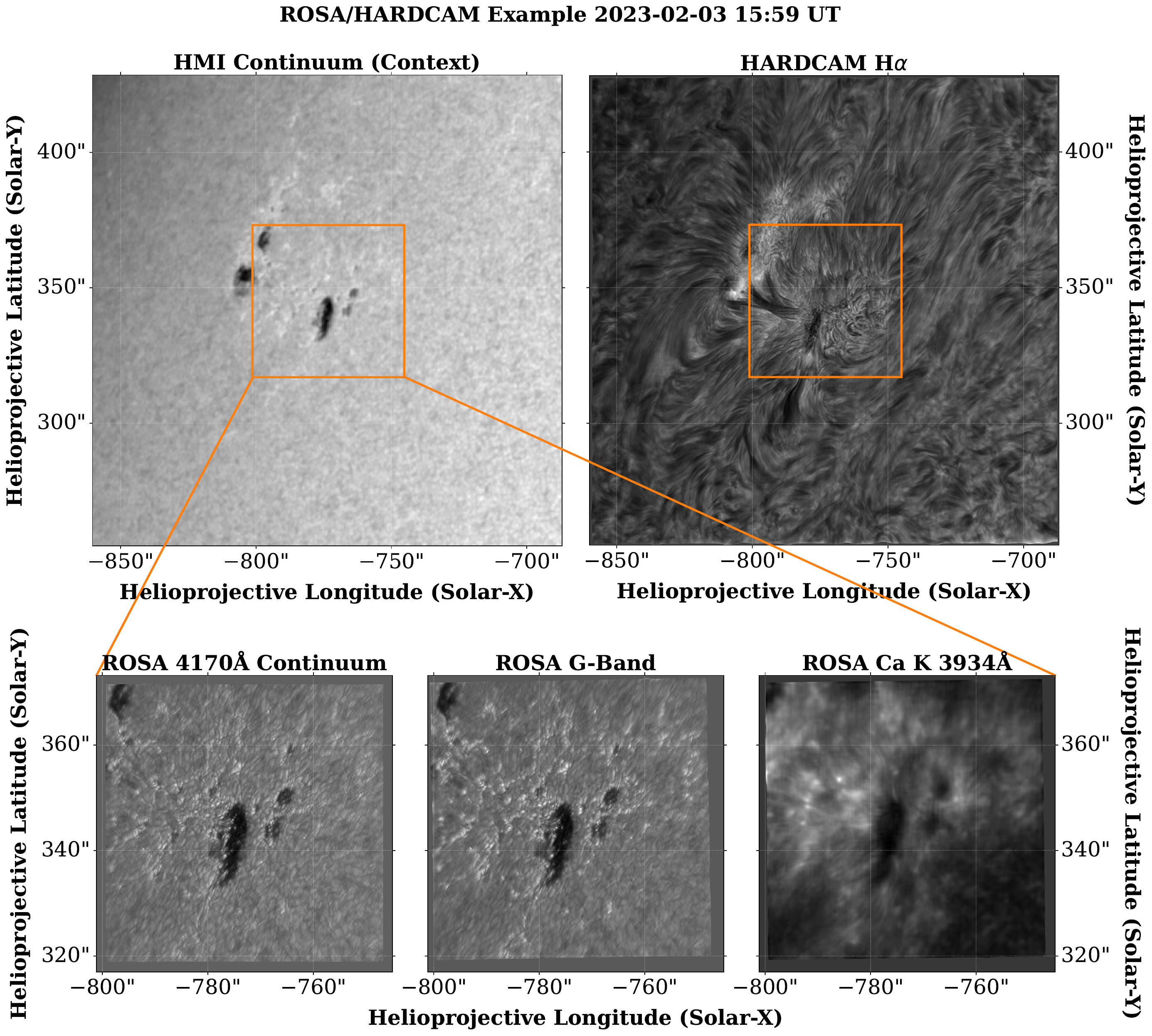}
    \caption{Speckle-reconstructed images from the ROSA (bottom row, left-to-right are images of the 4170~{\AA} continuum, G-Band, and Ca~K line core) and HARDcam (top right) instruments, compared to the cotemporal HMI continuum image (top left). The images chosen for display represent the mean seeing conditions and image reconstruction capabilities available within the SSODA. The orange square displayed on the HMI and HARDcam images denotes the FOV observed by the ROSA cameras. The diffraction limit of the DST is between $0{\,}.{\!\!}''116 - 0{\,}.{\!\!}''142$ for the ROSA system (3500~{\AA} $-$ 4300~{\AA}), and is $0{\,}.{\!\!}''217$ at 6563~{\AA}. For this observing series, the ROSA cameras were sampled at $0{\,}.{\!\!}''056$ per pixel, while HARDcam was sampled at $0{\,}.{\!\!}''0846$ per pixel.}
    \label{fig:rosa-ex}
\end{figure}

Examples of speckle-reconstructed and destretched ROSA/HARDcam data are displayed in Figure~\ref{fig:rosa-ex}, which shows examples of the system operating within the diffraction-limited regime. 
The data selected for display were acquired on February 3, 2023, which is representative of an observing run with average seeing conditions. 
Further, the selected frames were reconstructed over a period corresponding to the approximate mean seeing during observations on that date. 

\subsection{SPectropolarimeter for INfrared and Optical Regions}\label{sec:instr.spinor}
\begin{figure}
    \centering
    \includegraphics[width=\textwidth]{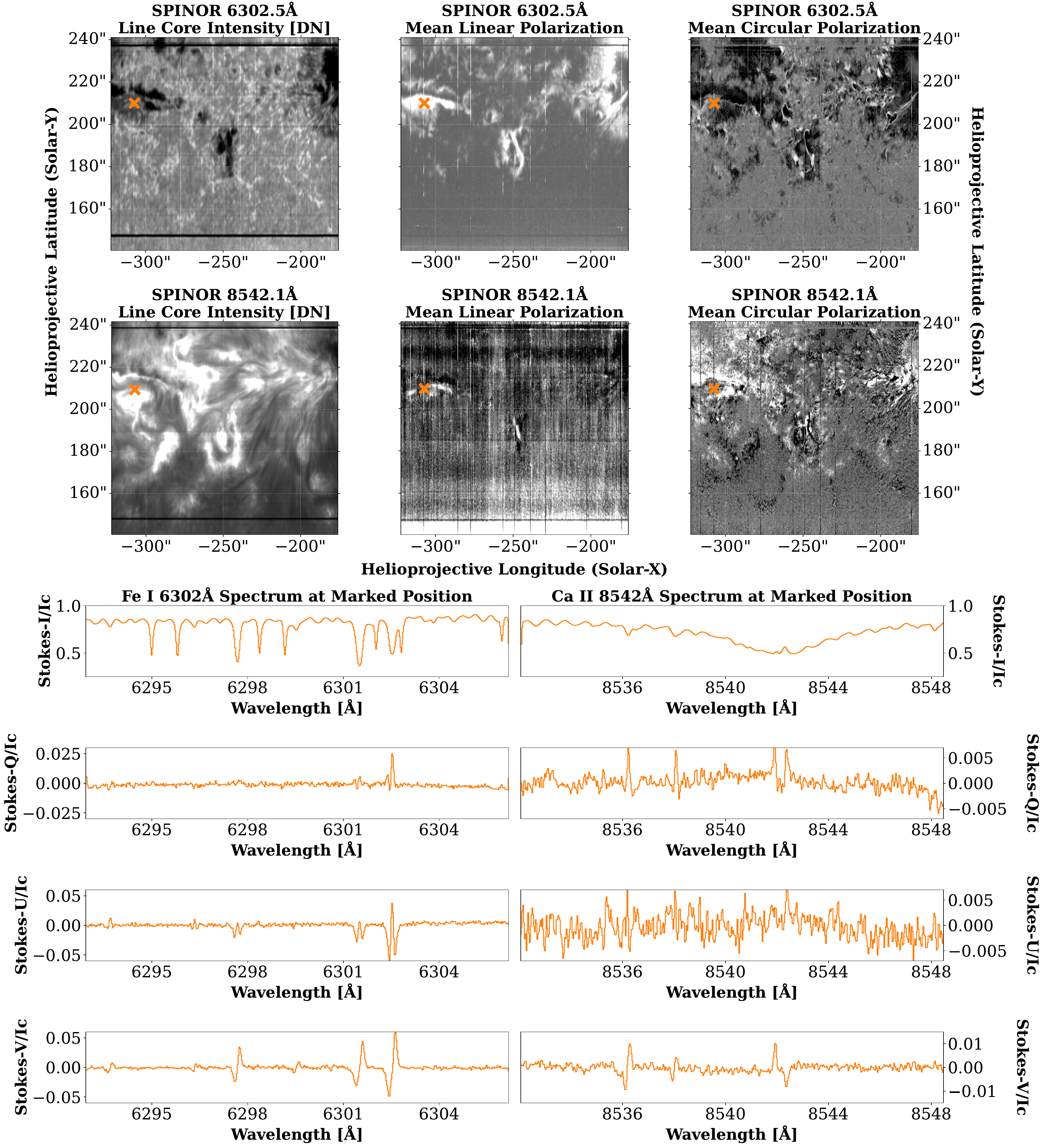}
    \caption{Example of SPINOR Level-1.5 data from February 3, 2025. On this date, the Fe~{\sc{i}}~6302~{\AA} and Ca~{\sc{ii}}~8542~{\AA} spectral windows were observed. The top row shows, from left-to-right, the line core intensity of Fe~{\sc{i}}~6302~{\AA}, the mean linear polarization of the line, and the mean circular polarization of the line, using methods described in \citet{2011SoPh..268...57M}. The second row shows these same products for the Ca~{\sc{ii}}~8542~{\AA} line. The bottom half of the figure shows the reduced Stokes profiles for the position marked by an orange ``x'' in the top two rows. The left column displays the Stokes profiles for Fe~{\sc{i}}~6302~{\AA} at this position, while the right column displays the Ca~{\sc{ii}}~8542~{\AA} profiles.}
    \label{fig:spinor-ex}
\end{figure}

The SPectropolarimeter for Infrared and Optical Regions \citep[SPINOR;][]{SPINOR} is a dual-beam rastering slit spectropolarimeter with cotemporal slit-jaw imaging. 
The instrument can be configured to offer spectropolarimetry for a variety of optical and near-infrared lines using up to four synchronized cameras.
During consortium operations, the chromospheric Ca~{\sc{ii}}~8542~{\AA}, photospheric Fe~{\sc{i}}~6302~{\AA}, and chromospheric He~{\sc{i}}~D$_{3}$~5876~{\AA} lines are typically selected for study, with slit-jaw images acquired in a broadband 6302~{\AA} filter, with a 4~{\AA} bandpass.
During normal consortium operations, the instrument is operated with an $0{\,}.{\!\!}''3$ slit width, with a plate scale of $0{\,}.{\!\!}''15 - 0{\,}.{\!\!}''20$ per pixel along the $\approx 90${\arcsec} slit.
When operated in a single-beam configuration without polarimetric capabilities, the instrument is referred to as the Horizontal SpectroGraph (HSG), and is capable of faster data acquisition.
SPINOR data are available within the SSODA as Level-0, Level-1, and Level-1.5 data products. These data levels correspond to the following data products:
\begin{itemize}
    \item \textbf{Level-0:} SPINOR Level-0 data are the raw data obtained from daily observations. 
    These data are stored in FITS format, with each file corresponding to a single raster scan. 
    Within each file, each extension corresponds to a single raster position, with eight polarization modulation states within each extension.
    \item \textbf{Level-1:} SPINOR data at level one are reduced via the legacy IDL pipeline, largely written by Dr. Christian Beck. 
    This pipeline corrects for dark current, flat-fielding, and performs a polarimetric calibration, including \textit{I} $\rightarrow$ \textit{Q}, \textit{U}, \textit{V} crosstalk. 
    SPINOR data at level one are stored as binary, with an IDL save file providing metadata from the calibration process. 
    \item \textbf{Level-1.5:} The SPINOR Level-1.5 data product is a new data product designed to make best use of a new polarimetric modulator, installed in September 2024. Level-1.5 data are repackaged as FITS format, and include a similar correction sequence as the Level-1 data product with optional corrections for \textit{V} $\leftrightarrow$ \textit{QU} crosstalk.
\end{itemize}
While the SPINOR pipeline does not currently offer a Level-2 data product, this product is under active development, and is anticipated to be part of future data releases. 
An example of SPINOR data can be seen in Figure~\ref{fig:spinor-ex} from the date Febraury 3, 2025. On this date, SPINOR was operated with full-Stokes spectropolarimetry of the Fe~{\sc{i}}~6302~{\AA} and Ca~{\sc{ii}}~8542~{\AA} line complexes. 
Maps of the line core intensity and the mean degree of linear and circular polarization are shown for these lines in the top half of the figure, while the bottom half of the figure displays Stokes profiles extracted from the location marked in orange on each of the panels of the upper half of the figure.


\subsection{Horizontal Spectro-Graph}\label{sec:instr.hsg}
The Horizontal Spectro-Graph (HSG) is a configurable rastering slit spectrograph with cotemporal slit-jaw imaging. During consortium operations, typical spectral windows include Fe~{\sc{i}}~6302~{\AA}, He~{\sc{i}}~5876~{\AA}, Ca~{\sc{ii}}~8542~{\AA}, and Na~{\sc{i}}~5896~{\AA}. 
Slit-jaw images are typically acquired using a broadband 6302~{\AA} filter with a 4~{\AA} bandpass. 
Since 2023, during active region observations, the instrument is frequently operated in a fast-rastering mode, with 16 or 32 slit positions separated by either $0{\,}.{\!\!}''3$ or $0{\,}.{\!\!}''6$. 
This results in a raster cadence between 25~seconds and one minute, depending on the specific setup chosen. 
HSG data are available within the SSODA as Level-0, Level-1, and Level-1.5 data products. 
These data levels correspond to the following data products:
\begin{itemize}
    \item \textbf{Level-0:} HSG Level-0 data are the raw data obtained from daily observations. 
    These data are stored in FITS format, with each file corresponding to a single raster scan. 
    Within each file, each extension corresponds to a single slit position.
    \item \textbf{Level-1:} HSG Level-1 data are the minimally reduced data product. Data at Level-1 have been clipped to the beam size, and gain corrected. Level-1 data products are stored in IDL save format and constructed via a partial implementation of the \texttt{firs-soft} Level-1 pipeline. This data product is considered obsolete, and has largely been replaced by the Level-1.5 data product.
    \item \textbf{Level-1.5}: HSG data at Level-1.5 are reduced using the hsgPy subpackage, available as part of the SSOsoft calibration package. Data at Level-1.5 have been clipped to beam size, gain corrected, wavelength calibrated, and optionally subjected to spectral transmission and fringe corrections. Level-1.5 data are also subjected to moment analysis in user-defined spectral ranges. Level-1.5 data are stored in FITS format, with the zeroth extension containing the reduced spectral data and subsequent extensions containing integrated intensity, LOS velocity, and velocity width from moment analysis. The final extension contains observational metadata, including the wavelength grid, time elapsed from raster start, exposure time per slit position, and telescope parameters for latitude, longitude, rotation from solar-north, light level, and Seykora scintillation monitor values. 
\end{itemize}

\subsection{Interferometric BI-dimensional Spectropolarimeter}\label{sec:instr.ibis}
\begin{figure}
    \centering
    \includegraphics[width=0.95\textwidth]{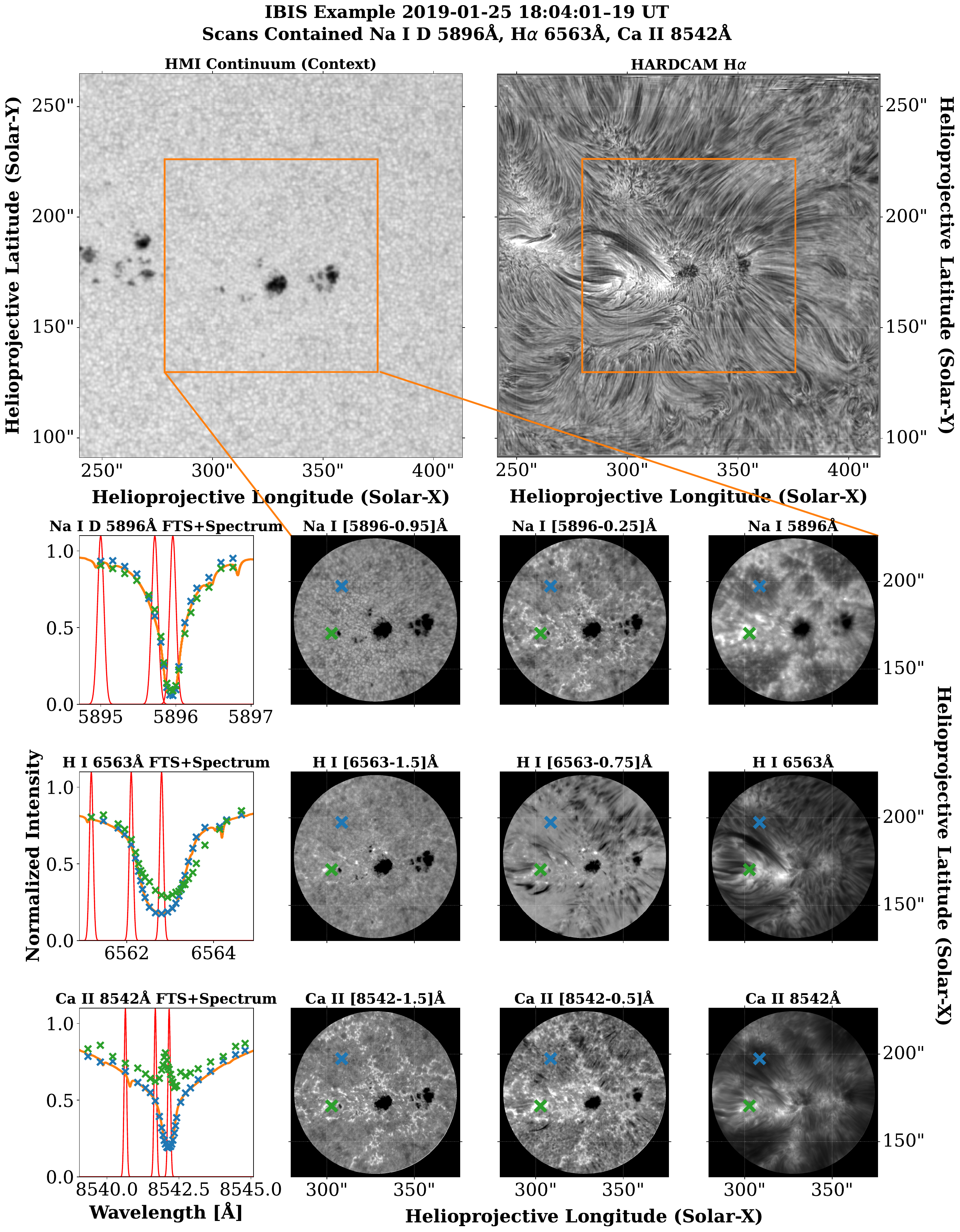}
    \caption{Example of IBIS data from January 25, 2019. The top row shows the wider region in the HMI continuum (top left) and HARDcam H$\alpha$ (top right), while successive rows show the line profile, far wing, near wing, and core of the Na~{\sc{i}}, H~{\sc{i}}, and Ca~{\sc{ii}} lines respectively. The H$\alpha$ line-core images from IBIS (0.1~{\AA} FWHM) and HARDcam (0.25~{\AA} FWHM) are similar, but not exactly the same, due to the broader bandpass of HARDcam, the images from which have also been subjected to speckle reconstruction techniques. The black circle in each IBIS image is due to a circular 90\arcsec\ field stop within the IBIS instrument, which masks edge regions that can be susceptible to significant blueshifts. The orange line profiles in the left column are from the FTS atlas. The red profiles overlaid are representative of the IBIS filter function at the three spectral positions displayed in the right three columns. The blue and green overlaid marks are the IBIS spectral profiles at the positions marked in the same color on the narrowband filter images. The IBIS data from this date are representative of a fine-sampling mode, with many spectral positions in each scan.}
    \label{fig:ibis-ex}
\end{figure}

The Interferometric BI-dimensional Spectropolarimeter \cite[IBIS;][]{IBIS1, IBIS2, IBIS3} was a dual Fabry-P\'erot scanning spectropolarimeter, operated under consortium direction from $2017-2019$. 
The instrument ceased observations on June 27, 2019, after more than a decade of operation as a facility instrument.
The instrument concept was designed to provide scans of individual spectral lines, taking a series of discrete images at several wavelength positions across the line profile. 
A cotemporal broadband imager provided a reference channel for the purposes of rigid image destretching. 
The IBIS instrument typically operated with a 90\arcsec\ circular field-of-view, and could, upon the investigator's request, be operated with full-Stokes polarimetry.
A variety of spectral lines could be selected for study, with daily observations typically including either the H$\alpha$~6562.8~{\AA} or Ca~{\sc{ii}}~8542~{\AA} line. 
IBIS data are available within the SSODA as Level-0 and Level-1 data products. 
These data levels correspond to the following data products:
\begin{itemize}
    \item \textbf{Level-0:} IBIS Level-0 data are the raw data obtained from daily observations. 
    These data are stored in extended FITS format. 
    Each FITS file corresponds to one full wavelength sweep of the given subset of spectral lines, with the number of repetitions determined by the observer, and each extension within the FITS file corresponds to a single wavelength position.
    \item \textbf{Level-1:} IBIS data are reduced via the standard IDL pipeline. 
    This pipeline corrects for dark current, flat-fielding, and blueshifts across the FOV, then destretches the narrowband data relative to the simultaneous broadband channel. 
    Level-1 data are stored either as IDL save format, or are repacked to FITS for convenience.
\end{itemize}
A separate archive of IBIS data, including other derived quantities, is maintained by the Istituto Nazionale di Astrofisica (INAF). 
Details on this archive can be found in \cite{IBIS-A}. 
An example of SSODA IBIS Level-1 data is displayed in Figure~\ref{fig:ibis-ex}. 
Figure~\ref{fig:ibis-ex} contains data from January 25, 2019. On this date, IBIS observed three spectral lines, the Na~{\sc{i}}~D~5896~{\AA} line (19 spectral positions), the H$\alpha$~6563~{\AA} line (27 spectral positions), and the Ca~{\sc{ii}}~8542~{\AA} line (30 spectral positions). 
Selections from the far blue wing, near blue wing, and line core are displayed in Figure~\ref{fig:ibis-ex}. 
The left column in each row contains the line profile obtained from the FTS atlas as the orange solid line. Green and blue ``x'' symbols denote the centroids of each IBIS scan position, corresponding to the profiles taken from the positions in the FOV marked by the same color. Red profiles show the IBIS filter profile at the scan positions chosen for display to the right. 
Near-wing positions were selected to display a mix of core and far-wing structures. Note that it is also possible to employ machine learning codes, such as those put forward by \citet{2021RSPTA.37900171M}, to further segregate different velocity components contained within each IBIS scan, which may help to unveil the multiple spectral signatures that contribute to the overall line profile shape. 

\section{Observing Campaigns}\label{sec:obs_camp}

\begin{figure}
    \centering
    \includegraphics[width=\textwidth]{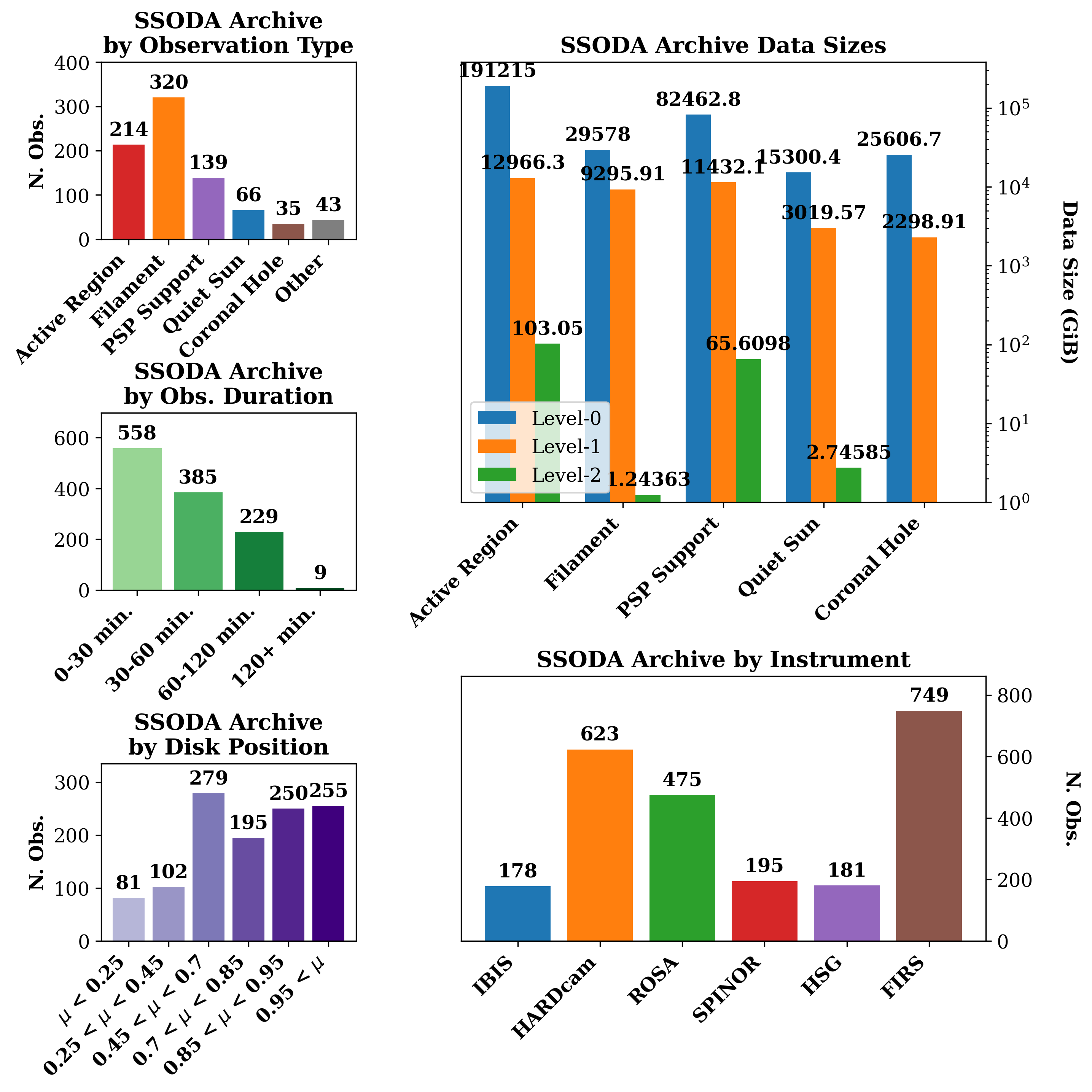}
    \caption{Overview of the SSODA as of April 24, 2025 by observing type, data volume, duration of science observations, disk position, and instrumental complement. The full SSODA on this date constituted 374~TiB of data (336 TiB at Level-0, 38~TiB at Level-1 or 1.5, and 170~GiB at Level-2), over 550 observing days, with a combined $\approx$683 hours of observations at 1162 discrete coordinate positions.}
    \label{fig:overview}
\end{figure}

In addition to PI-led operations, and targets of opportunity, the DST under consortium operations has five primary synoptic observing modes:
\begin{itemize}
    \item \textbf{Flare \& Active Region Studies:} In flare-focused active region configuration, the DST is operated covering major active regions in a variety of photospheric and chromospheric passbands. This observing mode typically uses all available instrumentation.
    \item \textbf{Filament Studies:} During filament observation campaigns, quiet-Sun filaments were prioritized, though many active region filament datasets exist as well. In the filament observation configuration, FIRS data are prioritized. Filament observations typically include FIRS spectropolarimetric maps of on-disk filaments in addition to context imaging from a chromospheric channel, such as HARDcam, or IBIS (until 2019). For examples of DST filament-mode observations and significant results, see \cite{Shuo2020, Shuo2022}.
    \item \textbf{Parker Solar Probe (PSP) Footpoints:} In PSP co-observing mode, the DST is operated to cover regions identified as the footpoints of magnetic structures connected to the Parker Solar Probe, typically during the probe's perihelion. DST observations during PSP-related campaigns are frequently near the solar limb, and use a subset of instrumentation. The most frequent instrumental configuration includes the ROSA Ca~K and G-band channels, HARDcam, FIRS 10830~{\AA}, and HSG maps of the Fe~{\sc{i}}~6302~{\AA}, Ca~{\sc{ii}}~8542~{\AA}, and Na~{\sc{i}}~D~5893~{\AA} lines.
    \item \textbf{Quiet Sun Studies:} Quiet Sun observations are typically made near disk center ($76\%$ of datasets taken under this observing mode are at a $\mu > 0.95$), and are made at or near the telescope's diffraction limit for a given instrument, with all available instrumentation operating, providing high-cadence imaging, spectroscopy, and spectropolarimetry of quiescent granulation. 
    \item \textbf{Coronal Hole Studies:} The instrumentation setup used during PSP support campaigns has also been used to capture bright point motions along coronal hole boundaries. These datasets are high-cadence, but typically use a lower spatial resolution of the ROSA cameras in order to expand the FOV of G-Band and Ca~K channels to 150\arcsec$\times$150\arcsec .  Coronal hole datasets frequently overlap with PSP support datasets.
\end{itemize}

Figure~\ref{fig:overview} shows a top-level breakdown of data within the archive sorted by data level, observation type, duration, instrument use, and position on the solar disk, $\mu$, where $\mu$ is the cosine of the heliocentric angle. 
FIRS is the most frequently operated instrument at the DST, used in nearly every observing campaign for its ability to obtain high-quality spectropolarimetric observations of the photosphere and chromosphere. 
While filament observation campaigns are the most numerous, they have a comparatively low data volume, due to the reduced instrumental complement and cadence used during filament observations. 
The typical observation within the archive is between $60$~and~$120$~min. in duration (not including time reserved for calibrations), with 41\% of observations within this duration. 
Observations less than 60~min. in duration comprise another 42\%, and observations of an extended duration comprise the final 17\%. 
Most observations occur near the center of the solar disk. 
Observations with $\mu > 0.85$, corresponding to the approximate disk center, comprise 50\% of the archive. 
Intermediate disk positions, with $0.45 < \mu < 0.85$ comprise a further 36\%, and observations at, near, or above the solar limb ($\mu < 0.45$) form the last 14\%. 
The observations of the latter group are comparatively few due primarily to the difficulties of operating a solar adaptive optics system at the solar limb, where higher-order atmospheric corrections may be unavailable due to the decrease in structural contrast caused by limb darkening.

\section{Data Quality}\label{sec:quality}

Adequate tracking of data quality is essential to providing a user-friendly final product. To this end, several parameters are tracked for each observing series in order to allow the user to select for datasets that suit their particular science use case. The SSODA tracks the mean scintillation measured during a given observing series, the duration of the observing series, and a selection of events of interest, determined by querying the Heliophysics Events Knowledgebase (HEK).

\subsection{DST Seeing Quality}
In the case of ground-based solar data, determining seeing quality is nontrivial. 
The Dunn Solar Telescope is equipped with a Seykora scintillation monitor \citep{seykora}, which provides a real-time measure of atmospheric scintillation.
The average of these measurements is used to characterize the seeing quality over a given observing series. Table~\ref{tab:qual} shows the best-seeing datasets for four of the DST synoptic modes. Scintillation, however, is perhaps better understood as a first-order estimate of seeing quality.
Seykora scintillation values are insensitive to upper atmospheric perturbances (such as those caused by the jet stream), and are decoupled from the effects of adaptive optics.
As solar adaptive optic systems operate via cross-correlation of a number of subapertures, the quality of AO ``lock'' may vary between different structures, with compact, high-contrast features (such as small sunspots) providing the highest-quality of lock with the least attention required from telescope operators. Extended sources (e.g., large sunspots) and low-contrast sources (e.g., quiet Sun granulation) require significantly more effort to maintain high-quality atmospheric corrections.
\par
In order for the user to understand the seeing properties of a given dataset, metrics sensitive to the quality of AO-compensated seeing are required. To this end, every observing series acquired with any of the ROSA or HARDcam cameras has been parameterized with a variety of image-quality metrics.
Of the various available image quality metrics, two have been selected as the primary metrics for determining seeing quality: the Helmli-Scherer mean \citep[HSM;][]{HSM} for ROSA data, and the Median Filter Gradient Similarity metric \citep[MFGS;][]{MFGS} for HARDcam data.
\citet{ImageQuality} found that these methods are extremely robust as stand-in metrics for image quality. We have found that HSM can be unreliable for chromospheric structures, such as those observed by HARDcam, hence the selection of MFGS for this camera. HSM is retained for ROSA cameras, primarily because it is significantly faster to compute.
\par
\begin{figure}
    \centering
    \includegraphics[width=0.94\textwidth]{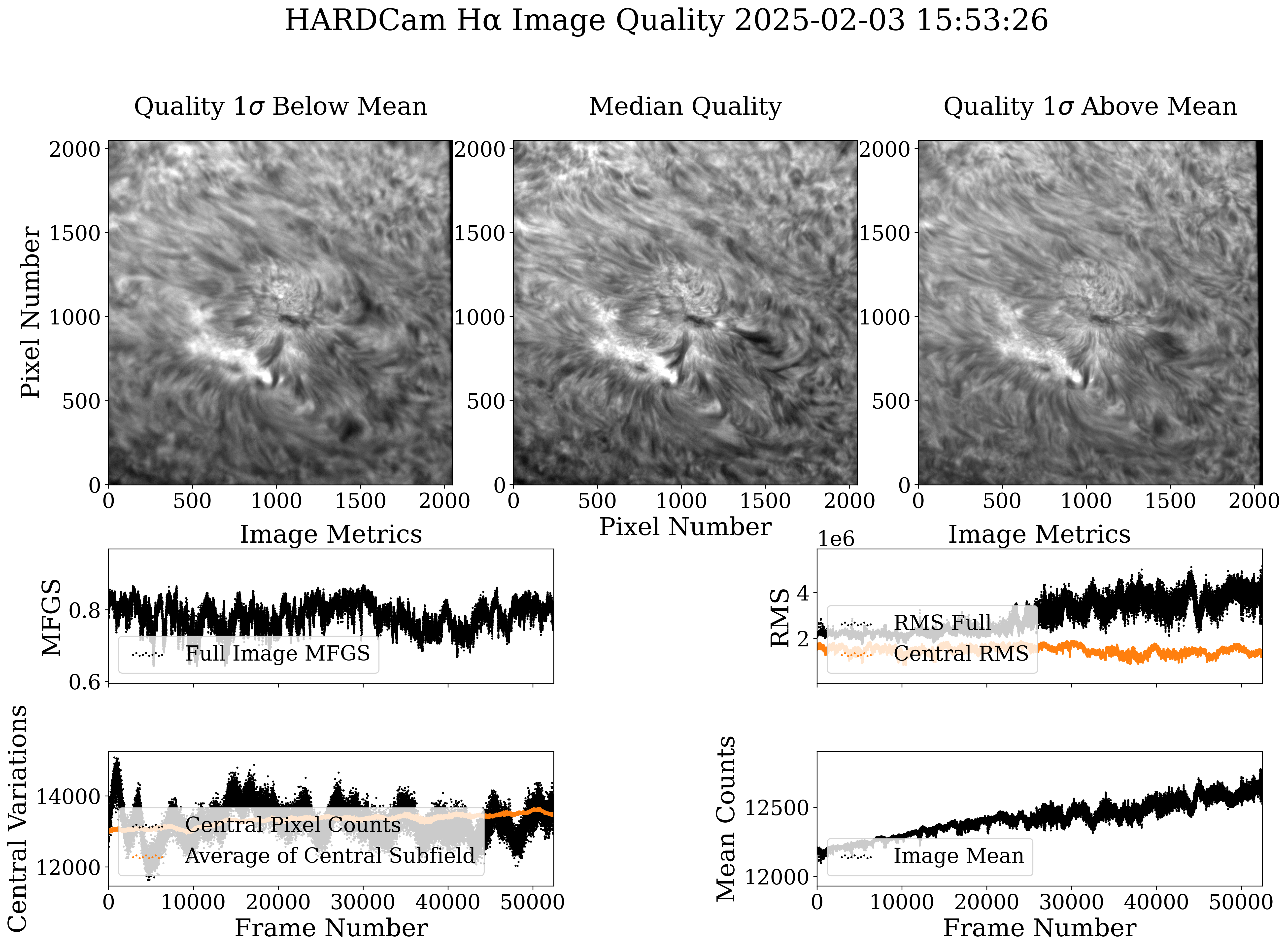}
    \includegraphics[width=0.94\textwidth]{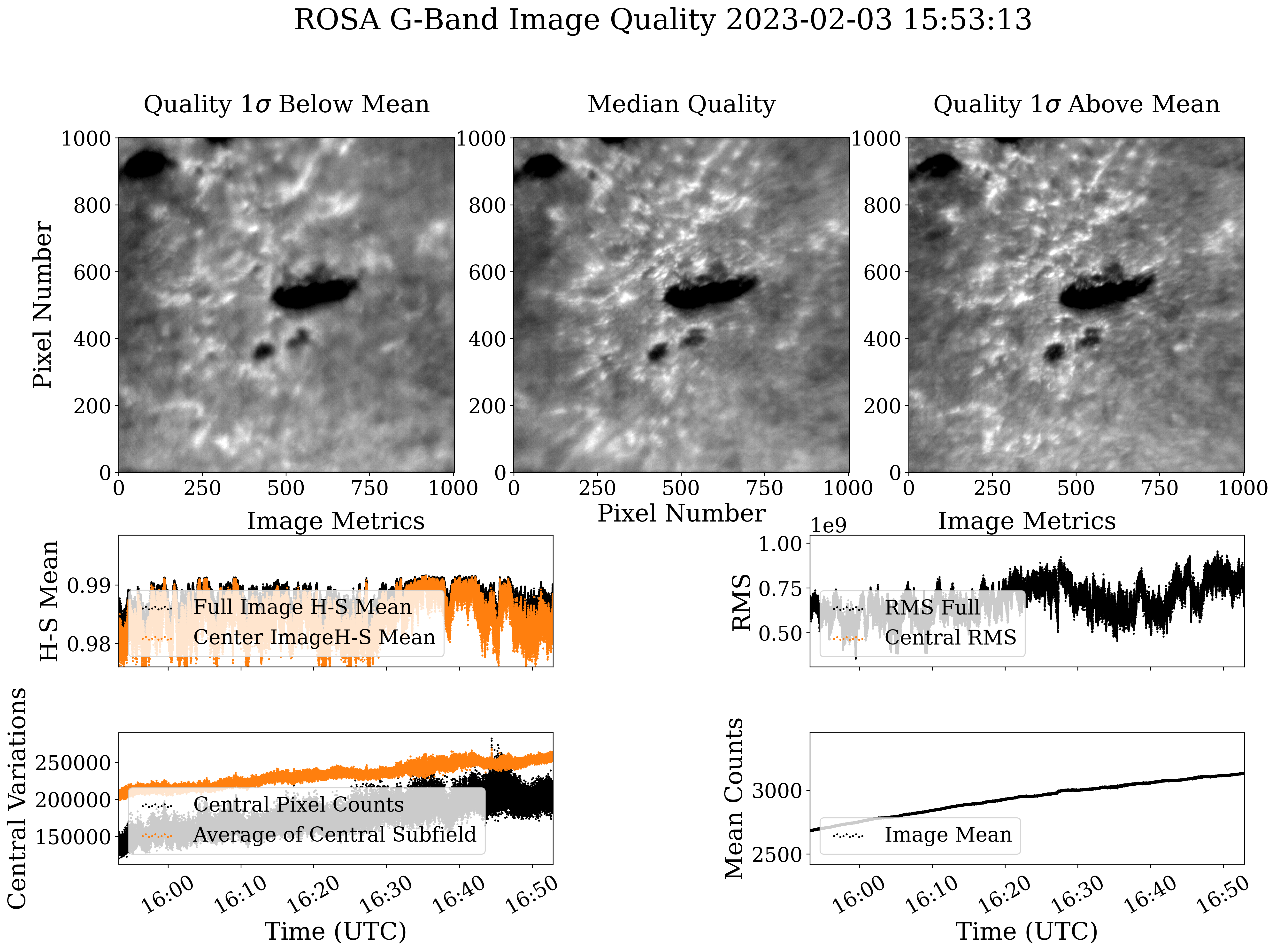}
    \caption{Automatically-generated overview seeing plots from February 3, 2023, indicative of average seeing quality at the DST. HARDcam data are shown in the top half of the figure, while ROSA G-band data are shown below. For each plot, the top row shows, left-to-right, an image representative of seeing at: one standard deviation worse than the mean, the median seeing quality, and one standard deviation better than the mean. The four panels below show the computed image quality metrics. The top left shows the HSM or MFGS (depending on the camera), and proceeding clockwise, the root-mean-squared (RMS) value, the image mean, and the mean of the central subfield are shown for the entire observing sequence.}
    \label{fig:seeing}
\end{figure}

For every observing day with imaging data, the selected metrics were calculated frame-by-frame from Level-0 data. Level-1 data are not used for this purpose due to the application of speckle reconstruction, which will nearly always produce an image that is ``perfect'' to image quality metrics, having significant contrast on small spatial scales, however, in poor seeing conditions, this contrast can be induced by the reconstruction process.
These metrics are presented in full on the summary page for each date within the archive, along with representative Level-0 images taking during a period of median seeing, as well as images bracketing the $\pm 1\sigma$ surrounding the mean quality. An example of this presentation can be found in Figure~\ref{fig:seeing}, for ROSA G-band and HARDcam H$\alpha$ filtergrams. Figure~\ref{fig:seeing} shows the seeing plots calculated from data obtained on February 3, 2023, which is representative of an observing day with average seeing. Figure~\ref{fig:rosa-ex} shows the reconstructed images (Level-1.5) corresponding to the median-seeing images in Figure~\ref{fig:seeing}.

\par

\subsection{\textbf{Tracked Events}}
Once seeing is parameterized, the telescope pointing and observation timing are checked against a subset of events from the HEK database. Currently, solar flares, filament eruptions, and coronal rain (as some observations extend beyond the limb) events are tracked, along with the intersection of the DST FOV and the IRIS FOV. Currently, an event is considered as valid within the SSODA if the the DST coverage of the event is at least partial. For the purposes of the SSODA, partial coverage is defined as an event occurring either at the edge of the DST FOV with the full time span covered, or occurring at the center of the DST FOV, but without full temporal coverage of the event (e.g., capturing only the decay phase of a flare).

\begin{table}[]
    \centering
    \begin{tabular}{lcccccc}
        \hline
        \multicolumn{7}{c}{\textbf{Filament Observations with High-Quality Seeing}} \\
        \hline 
        Date & Best Scin. & Duration  & Mean Scin.  & Full Day & Events & Position \\
             & [arcsec]   & [min]   & [arcsec] & Dur. [min] &        & $\mu$ \\
        \hline
        2020-11-03 & 0.40 & 41 & 0.72 & 210 &  & 0.40 \textendash 0.53 \\
        2021-02-25 & 0.42 & 41 & 1.5  & 133 &  & 0.60 \textendash 0.72 \\
        2018-09-24 & 0.46 & 48 & 0.65 & 83  &  & 0.72 \textendash 0.73 \\
        2019-03-19 & 0.47 & 33 & 0.55 & 66  &  & 0.94 \textendash 0.95 \\
        2020-07-30 & 0.63 & 61 & \textendash & 61 &  & 0.71 \\
        \hline
        \hline
        \multicolumn{7}{c}{\textbf{Active Region Observations with High-Quality Seeing}} \\
        \hline
        Date & Best Scin. & Duration  & Mean Scin.  & Full Day & Events & Position \\
             & [arcsec]   & [min]   & [arcsec] & Dur. [min] &        & $\mu$ \\
        \hline
        2024-11-22 & 0.36 & 32 & 0.90 & 141 & Sigmoid & 0.89 \textendash 0.97 \\
        2021-11-09 & 0.43 & 41 & 0.52 & 82 & Sigmoid & 0.79 \textendash 0.84 \\
        2023-09-20 & 0.44 & 61 & \textendash & 61 & M8.2 & 0.83 \\
        2023-05-19 & 0.45 & 51 & \textendash & 51 &  & 0.56 \\
        2025-03-31 & 0.67 & 72 & \textendash & 72 & C2.2 & 0.47 \\
        \hline
        \hline
        \multicolumn{7}{c}{\textbf{PSP Footpoint Co-Observing with High-Quality Seeing}} \\
        \hline
        Date & Best Scin. & Duration  & Mean Scin.  & Full Day & Events & Position \\
             & [arcsec]   & [min]   & [arcsec] & Dur. [min] &        & $\mu$ \\
        \hline
        2021-05-05 & 0.47 & 43 & 1.38 & 258 &  & 0.35 \textendash 0.49 \\
        2021-05-06 & 0.49 & 43 & 0.83 & 215 &  & 0.35 \textendash 0.50 \\
        2021-05-04 & 0.61 & 36 & 1.01 & 144 &  & 0.55 \textendash 0.63 \\
        2025-03-31 & 0.61 & 103 & 0.77 & 154 &  & 0.54 \textendash 0.96 \\
        2023-09-28 & 0.62 & 120 & \textendash & 120 &  & 0.01 \\
        \hline
        \hline
        \multicolumn{7}{c}{\textbf{Quiet-Sun Observations with High-Quality Seeing}} \\
        \hline
        Date & Best Scin. & Duration  & Mean Scin.  & Full Day & Events & Position \\
             & [arcsec]   & [min]   & [arcsec] & Dur. [min] &        & $\mu$ \\
        \hline
        2018-12-23 & 0.52 & 30 & 0.94 & 120 &  & 0.98 \\
        2022-08-01 & 0.53 & 58 & 0.57 & 116 & B9.7 & 1.0 \\
        2019-09-18 & 0.55 & 31 & 0.70 & 75 &  & 1.0 \\
        2019-09-23 & 0.60 & 32 & 0.75 & 63 &  & 1.0 \\
        2019-09-12 & 0.60 & 113 & \textendash & 113 & 1.0 \\
        \hline
  \end{tabular}
    \caption{Best-seeing datasets by observation type}
    \label{tab:qual}
\end{table}

\section{Data Access and Metadata}\label{sec:metadata}
Data taken under consortium operations are housed at the New Mexico State University Main Campus, and the archive can be accessed via the public portal\footnote{\url{http://ssoc.nmsu.edu}}. 
Certain observing series with oversized Level-0 data products may be housed in offline storage. 
In these cases, this is reflected on the daily observing series page, and the Level-0 data are made accessible on specific request.
\par
\begin{figure}
    \centering
    \includegraphics[width=\textwidth]{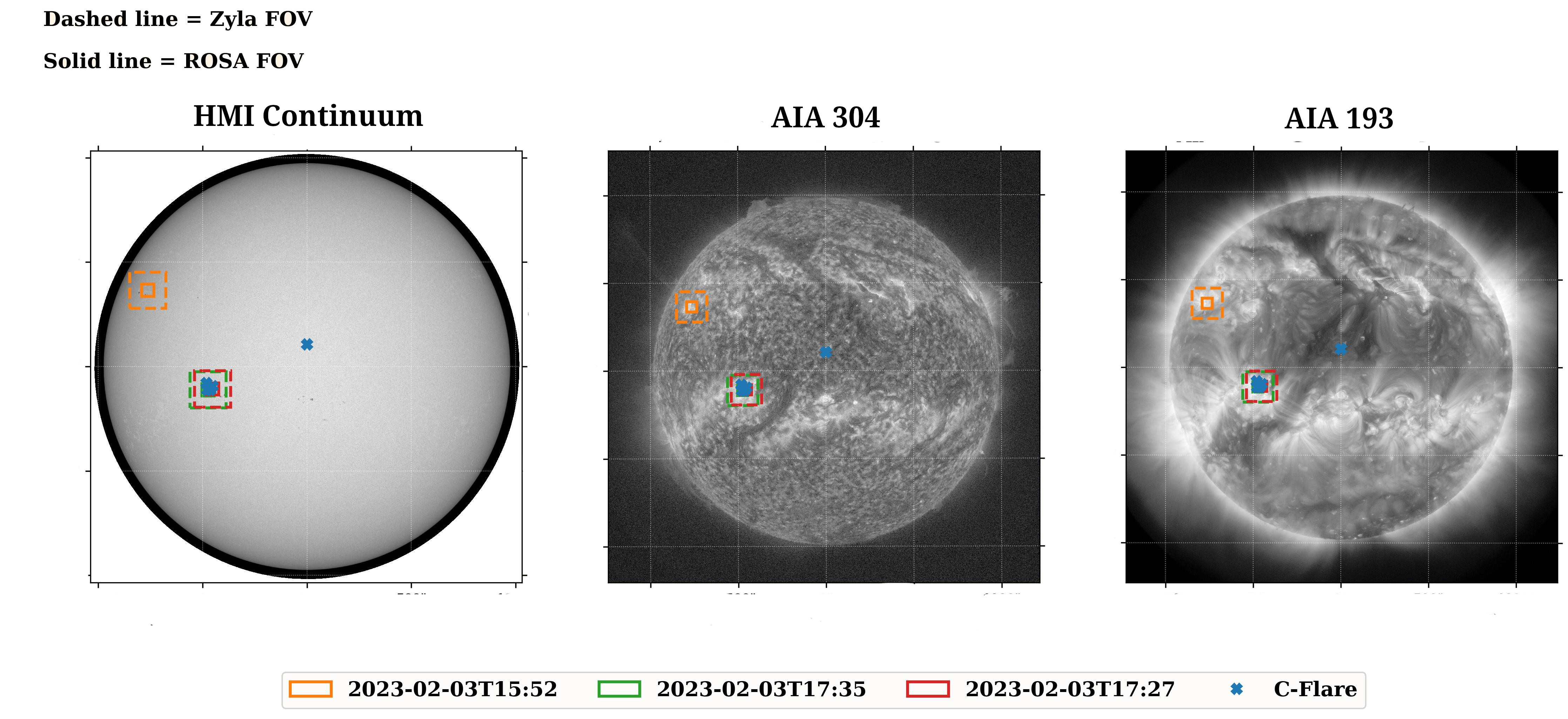}
    \includegraphics[width=0.85\textwidth]{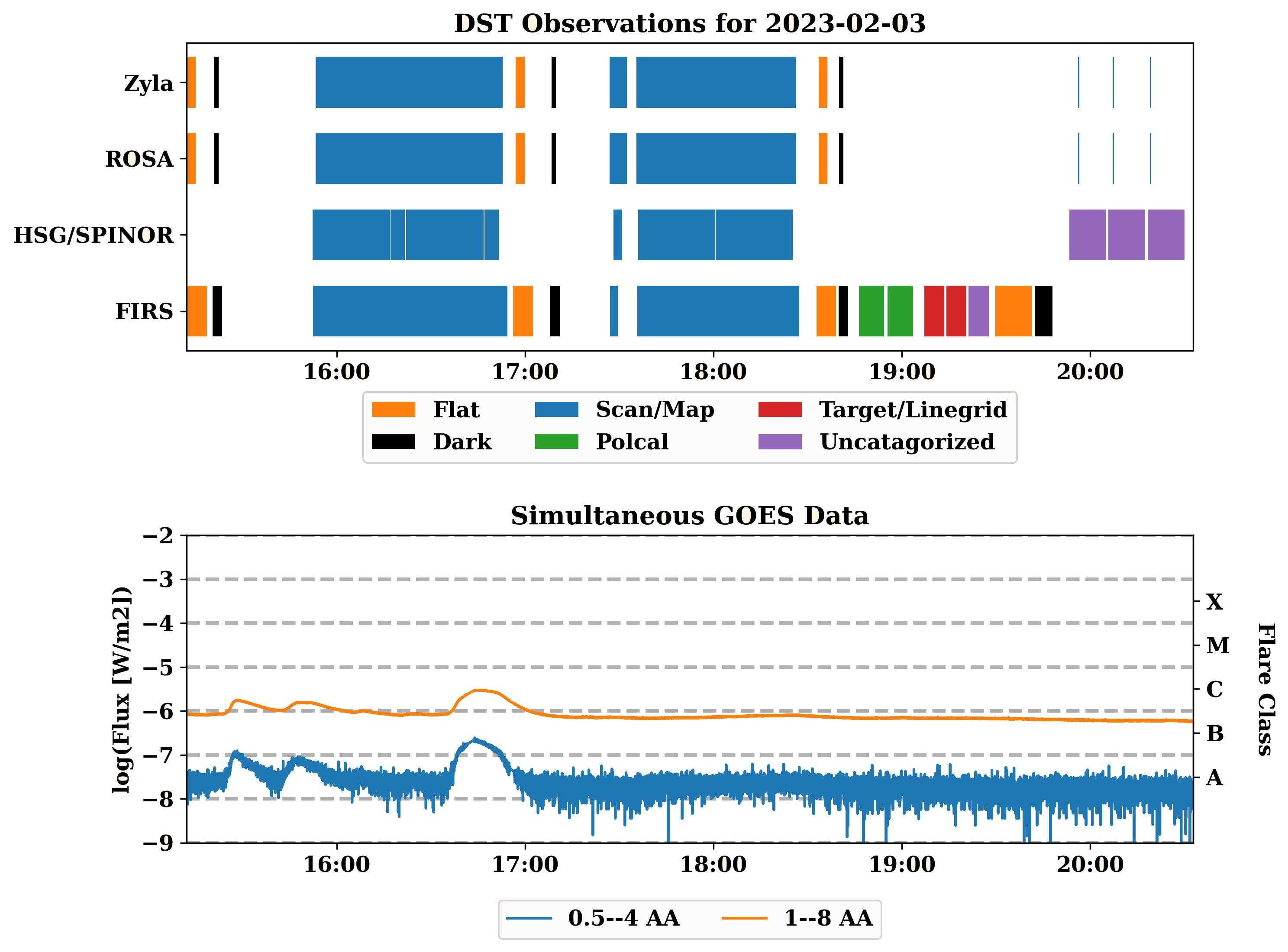}
    \includegraphics[width=0.55\textwidth]{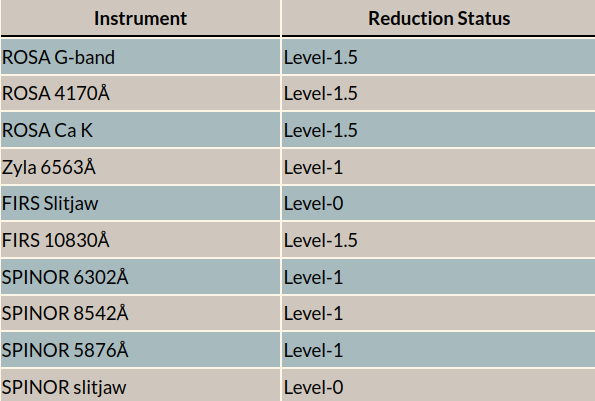}
    \caption{Collected metadata from the observing series on February 3, 2023. \textbf{Top:} Context imaging denoting the pointing and FOV of the DST. HEK events are overlaid. \textbf{Middle:} Observation summary plot denoting the timing of observations alongside GOES SXR flux for use in pinpointing flares. Both science and calibration observation durations are shown. \textbf{Bottom:} Table showing the reduction status of the instruments selected for daily observations.}
    \label{fig:metas}
\end{figure}

Within the primary archive, \path{/solardata}, data are organized datewise: \path{YYYY/MM/DD}.
The variety of file types present in the archive, numerous calibration series, and relatively small FOV observed by the telescope necessitate close tracking of observational metadata. 
This is collected and displayed on the landing page for each date within the archive, and is summarized on the landing page for each month. 
The overview pages are currently static HTML, which are updated as data proceeds through the reduction process. 
In addition to the previously-discussed seeing plots, these landing pages contain the telescope pointings for the day, alongside relevant HEK events, as well as the duration and timing of science and calibration observations, as compared to GOES X-ray data, and a table displaying the reduction status, all of which are automatically generated weekly upon data ingest, and updated as required. 
Examples of these metadata plots are displayed in Figure~\ref{fig:metas} from the daily overview page\footnote{\url{ssoc.nmsu.edu/solardata/2023/02/03}} for February 3, 2023.

During data ingest, the collected metadata parameters are also added to a searchable database\footnote{\url{http://ssoc.nmsu.edu/search/ObservationsDatabase/}}. 
The database is constructed using a SQLite backend, and managed through the Python-based Django framework. 
The observations database separates individual observing series and observing series metadata, of which there may be multiple over the course of an observing day. 
In addition to the information contained on the daily landing pages, the observations database tracks position on the disk, coronal hole overlap, NOAA Active Region Number, mean scintillation, and whether the IRIS FOV overlapped the telescope FOV during a series. 
Table~\ref{tab:qual} displays a summary of the five observing series with the best seeing available in the SSODA, separated by observation target. 
As part of the ongoing effort to provide streamlined data access, efforts to implement observing movies are underway.

\section{Science Use Case of the SSODA}\label{sec:usecase}

\begin{figure}
    \centering
    \includegraphics[width=\columnwidth]{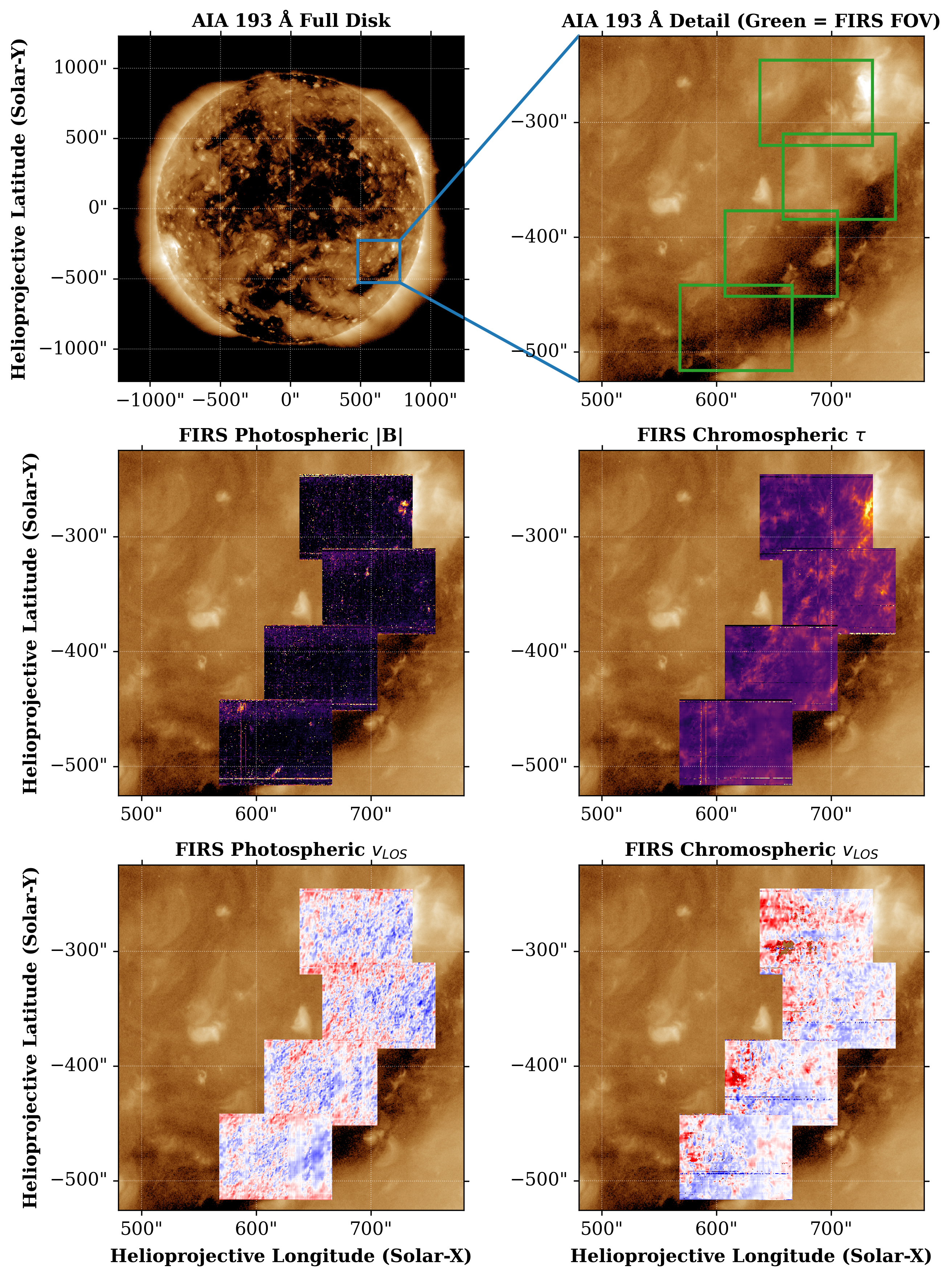}
    \caption{Mosaic of FIRS scan positions at a coronal hole boundary on May 4, 2021. \textbf{Top:} AIA 193~{\AA} context image of the full disk (left), and coronal hole region (right). The blue square in the full disk context image denotes the FOV shown in all other panels. \textbf{Middle:} Mosaic of FIRS maps along coronal hole boundary. The left panel shows the magnitude of magnetic field, $|\mathrm{\textbf{B}}|$, while the right panel shows the optical depth of chromospheric plasma, $\tau$. \textbf{Bottom:} LOS velocity mosaic for the photosphere (left) and chromosphere (right).}
    \label{fig:mosaic}
\end{figure}

\begin{figure}
    \centering
    \includegraphics[width=\columnwidth]{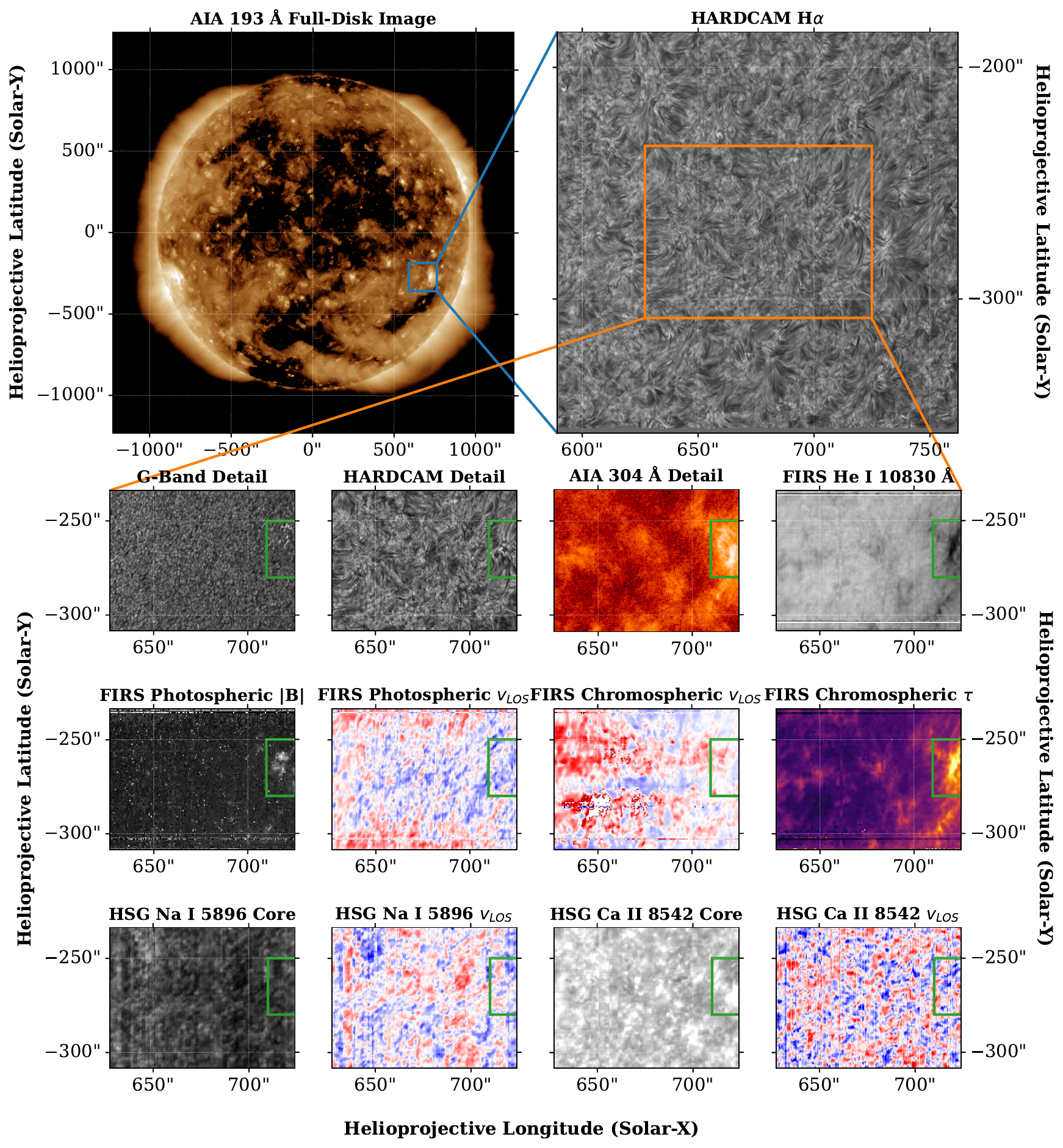}
    \caption{Data taken from an area of coronal brightening adjacent to a coronal hole on May 4, 2021. \textbf{Top:} AIA 193~{\AA} full-disk context image (left) and HARDcam H$\alpha$ wide-field image (right). \textbf{Second Row:} Region of interest (ROI) in G-band, H$\alpha$, AIA~304~{\AA}, and He~{\sc{i}} left-to-right. \textbf{Third Row:} ROI magnitude of photospheric magnetic field, photospheric LOS velocity, chromospheric LOS velocity, and optical depth of chromospheric plasma left-to-right. \textbf{Bottom:} ROI in HSG passbands for Na~{\sc{i}} line-core intensity and LOS velocity and Ca~{\sc{ii}} line-core intensity and LOS velocity left-to-right. Note that the velocities displayed in the HSG passbands are determined from moment analysis of the spectrum, and are not directly comparable to velocities obtained through spectropolarimetric inversions, such as those displayed for the FIRS data.}
    \label{fig:usecase}
\end{figure}

In this section, we showcase a simple example of DST data application from an observing campaign in May 2021. 
During this campaign, which lasted from May 1, 2021 to May 6, 2021, multiple spectral and spectropolarimetric scans were acquired daily along the boundary of a coronal hole as it progressed from near the center of the solar disk to the western limb. 
High cadence imaging was acquired in H$\alpha$, Ca~{\sc{ii}}~K, and G-band wavelengths, with spectroscopic observations of the Na~{\sc{i}}~D doublet, Ca~{\sc{ii}}~8542~{\AA} line, and the Fe~{\sc{i}}~6302~{\AA} complex, and spectropolarimetric observations of Si~{\sc{i}}~10827~{\AA} and He~{\sc{i}}~10830~{\AA}. 
All data shown in this section are publicly accessible through the SSODA portal.
\par
Figure~\ref{fig:mosaic} shows mosaics of FIRS Level-2 data products for four of the six maps acquired on May 4, 2021.
The final two maps were acquired during periods of poor seeing and have been omitted from this figure, but are still available through the SSODA.
All six maps were produced by spectropolarimetric inversions using the Hazel code \citep{hazel}, and were binned to have $0{\,}.{\!\!}''6$ square pixels to increase the polarimetric signals available in these weak field conditions.
These mosaics show that along the coronal hole boundary, more chromospheric material is present, and along this boundary structure, there are patches of enhanced photospheric magnetic field strength.
Sections of the boundary chromospheric structures also display upflows.
\par
Figure~\ref{fig:usecase} shows the other DST data products obtained for the northernmost FIRS raster. This raster contains, in part, a coronal brightening indicative of a jet adjacent to the coronal hole. DST data products can provide a cross-section of this structure through the photosphere and chromosphere. In this figure, the top row shows the location of the structure, both on the full solar disk and in the larger field. The H$\alpha$ image shows fibril structures cospatial with the coronal jet. The G-band image in the next row shows numerous brightenings indicative of the presence of magnetic bright points, and the image from the core of the He~{\sc{i}}~10830~{\AA} spectral line shows structure in the solar transition region. The following two rows show derived parameters, including magnetic field concentrations (from Si~{\sc{i}}~10827~{\AA}), and Doppler velocity maps from Si~{\sc{i}}~10827~{\AA}, He~{\sc{i}}~10830~{\AA}, Na~{\sc{i}}~5896~{\AA}, and Ca~{\sc{ii}}~8542~{\AA}. Doppler velocity maps are formed such that blue regions are indicative of motion towards the observer, while red regions are indicative of motion away from the observer. 
\par
From the Doppler velocity and magnetic field maps, the region cospatial with the structure visible in He~{\sc{i}} maps exhibits enhanced magnetic field, and the center of this structure shows upflows in Doppler velocity maps, while the periphery of this structure shows the return downflow. Available, but not shown are Doppler width maps for all observed spectral lines, and additional Doppler velocity maps for Na~{\sc{i}} and Ca~{\sc{ii}} lines. From an initial glance, these measurements reveal a structure originating in the solar photosphere and terminating in the corona with a variety of observations at each layer in the solar atmosphere. Tracking the evolution of the magnetic bright point structure, or performing time-distance analysis of the H$\alpha$ fibrils would provide further insight into this structure, and such analyses are facilitated by the SSODA.  

\section{Summary and Conclusions}\label{sec:conclusions}

We present the Sunspot Solar Observatory Data Archive, an ongoing effort to catalog and archive observations carried out by the Sunspot Solar Observatory Consortium using the Dunn Solar Telescope. The main archive contains over 375~TiB of data taken over more than 550 observing days. The SSODA is expected to grow as the DST remains in operation, covering the rise and peak of Solar Cycle 25, while introducing new instrumentation into its portfolio \citep[e.g., the FRANCIS integral field unit;][]{2023SoPh..298..146J}. This archive is intended to further scientific usage of the long-running observing campaigns carried out at the DST, which have applications to the studies of solar flares, quiet Sun and active region filament evolution, coronal hole dynamics, and coordination with PSP observations.
\par
In addition to raw data, the archive contains a significant amount of science-ready data, as well as higher-level data products containing derived parameters and inversion results. These data products include results of computationally-expensive procedures, including speckle-burst reconstruction and Hazel~2.0 inversion.
\par
As part of continuing efforts to provide high-quality, well-tagged data products, the SSODA includes searchable observational metadata, and charts of timing, position, and seeing, in order to efficiently select data for study. FITS headers include keywords that are partially compliant with standards like SOLARNET, and, in most cases, are compatible with high-level frameworks, such as Sunpy. 
\par
The SSODA provides a unique repository for high-quality observational solar data, ideal for studies of plasma processes, solar flares, and solar atmospheric dynamics. All archival data are publicly available, and the SSOC team is able to provide technical support and assistance in the best use of DST data. The data products served by the SSODA are under active development to better support the science use of the telescope. 

\begin{fundinginformation}
    J.S. acknowledges support from the National Science Foundations, grants 1936336 and 2401175. J.S. also acknowledges funding from the National Aeronautics and Space Administration, grant NNH23ZDA001N-BPSF.
    D.B.J. acknowledges support from the Leverhulme Trust via the Research Project Grant RPG-2019-371.
    D.B.J. wishes to thank the UK Science and Technology Facilities Council (STFC) for the consolidated grants ST/T00021X/1 and ST/X000923/1.
    D.B.J. also acknowledges funding from the UK Space Agency via the National Space Technology Programme (grant SSc-009).
    The ROSA instrument was supported by the UK STFC, while HARDcam observations are made possible by a Royal Society Research Grant (2009/R2). We also acknowledge partial support of this project from NASA grants 19-HSODS-004 and 21-SMDSS21-0047.
    We acknowledge that continuing operations at the DST are possible through generous support from a Research and Public Service Project funded by the State of New Mexico. We are also grateful for support from a subcontract between New Mexico State University and the Associated Universities for Research in Astronomy, with funding from the National Science Foundation.
\end{fundinginformation}

\begin{dataavailability}
    The data described in this work is housed at the SSODA portal, accessible at \url{ssoc.nmsu.edu}. The FIRS dataset selected as an example in Section~\ref{sec:instr.firs} can be accessed at \url{ssoc.nmsu.edu/solardata/2023/09}. The ROSA dataset selected as an example in Section~\ref{sec:instr.rosa} can be accessed at \url{ssoc.nmsu.edu/solardata/2023/02}. The SPINOR dataset selected as an example in Section~\ref{sec:instr.spinor} can be accessed at \url{ssoc.nmsu.edu/solardata/2025/02}. The IBIS dataset selected as an example in Section~\ref{sec:instr.ibis} can be accessed at \url{ssoc.nmsu.edu/solardata/2019/01}. The combined dataset displayed in Section~\ref{sec:usecase} can be accessed at \url{ssoc.nmsu.edu/solardata/2021/05}.
\end{dataavailability}

\begin{codeavailability}
    The original IDL-based reduction codes for the FIRS, SPINOR, and IBIS instruments are collected at \url{https://nso.edu/telescopes/dunn-solar-telescope/dst-pipelines/}.
    The FIRS and SPINOR IDL pipelines were written by Dr. Christian Beck, while the IBIS pipeline was written primarily by Drs. Serena Criscuoli and Alexandra Tritschler.
    \newline
    The current Python-based pipelines in use for ROSA, HARDcam, SPINOR, FIRS, and HSG data are available at \url{https://github.com/sgsellers/SSOsoft}. The original ROSA/HARDcam framework was written by Dr. Gordon MacDonald, and expanded to include FIRS, SPINOR, and HSG by Dr. Sean Sellers.
    \newline
    The original FIRS add-on pipeline for the creation on Level-1.5 and Level-2 data from the product of the IDL pipeline is available at \url{https://github.com/sgsellers/firs-tools}, however, it is considered depricated, and maintained only for backwards-compatibility.
    \newline
    FIRS Level-2 data products contain results of inversions carried out using the Hazel 2.0 code \citep{hazel}, which can be accessed at \url{https://aasensio.github.io/hazel2/}.
\end{codeavailability}

\begin{acks}
We thank the anonymous referee for their insightful comments and assistance in the final form of the manuscript, as well as the journal editors for formatting assistance. I thank Drs. R. Casini and D. Harrington for assistance with polarimetry calibrations and improvements to the SPINOR instrument and calibration software suggestions and improvements. I am grateful to Drs. C. Beck and K. Reardon for helpful discussions on instrumental calibration methods and equipment setup. I thank Dr. H. Lin for support and assistance with the FIRS instrument, as well as helpful discussions regarding polarimetric calibrations. I thank Dr. V. M. Pillet for his advocacy and assistance in the operation of the facility. I am grateful to Dr. O. Proch\'azka for improvements made to the archive structure. Finally, I thank D. Gilliam, for his valiant support of the DST spanning thirty years.


\end{acks}




\bibliographystyle{spr-mp-sola}
\bibliography{ssodabib} 

\end{article} 

\end{document}